\pdfoutput=1
\documentclass[12pt,a4paper]{article}

\usepackage{ifthen} 

\newboolean{pdflatex}
\setboolean{pdflatex}{true} 

\usepackage{lineno}

\newboolean{articletitles}
\setboolean{articletitles}{true} 

\newboolean{uprightparticles}
\setboolean{uprightparticles}{false} 

\def\paperauthors{LHCb collaboration} 
\def\paperasciititle{Three-pion Bose--Enstein correlations measured in proton-proton collisions} 
\def\papertitle{Three-pion Bose--Einstein correlations measured in proton-proton collisions} 
\def\paperkeywords{{High Energy Physics}, {LHCb}} 
\def\papercopyright{\the\year\ CERN for the benefit of the LHCb collaboration} 
\def\paperlicence{CC BY 4.0 licence}
\def\paperlicenceurl{https://creativecommons.org/licenses/by/4.0/}

\newif\ifEnableSectionTOCLinks
\EnableSectionTOCLinksfalse 

\usepackage[top=1in, bottom=1.25in, left=1in, right=1in]{geometry}

\usepackage{microtype}
\usepackage{lineno}  
\usepackage{xspace} 
\usepackage{caption} 

\usepackage{graphicx}  
\usepackage{color}
\usepackage{colortbl}
\graphicspath{{./figs/}} 

\usepackage{amsmath} 
\usepackage{amssymb}
\usepackage{amsfonts}
\usepackage{upgreek} 

\newcommand*\patchAmsMathEnvironmentForLineno[1]{%
\expandafter\let\csname old#1\expandafter\endcsname\csname #1\endcsname
\expandafter\let\csname oldend#1\expandafter\endcsname\csname
end#1\endcsname
 \renewenvironment{#1}%
   {\linenomath\csname old#1\endcsname}%
   {\csname oldend#1\endcsname\endlinenomath}%
}
\newcommand*\patchBothAmsMathEnvironmentsForLineno[1]{%
  \patchAmsMathEnvironmentForLineno{#1}%
  \patchAmsMathEnvironmentForLineno{#1*}%
}
\AtBeginDocument{%
\patchBothAmsMathEnvironmentsForLineno{equation}%
\patchBothAmsMathEnvironmentsForLineno{align}%
\patchBothAmsMathEnvironmentsForLineno{flalign}%
\patchBothAmsMathEnvironmentsForLineno{alignat}%
\patchBothAmsMathEnvironmentsForLineno{gather}%
\patchBothAmsMathEnvironmentsForLineno{multline}%
\patchBothAmsMathEnvironmentsForLineno{eqnarray}%
}

\usepackage[pdftex,
            pdfauthor={\paperauthors},
            pdftitle={\paperasciititle},
            pdfkeywords={\paperkeywords}]{hyperref}
\usepackage{hyperxmp}
\hypersetup{
    pdfcopyright={Copyright (C) \papercopyright},
    pdflicenseurl={\paperlicenceurl}
}

\usepackage[colorinlistoftodos,textsize=scriptsize]{todonotes}

\usepackage[bottom,flushmargin,hang,multiple]{footmisc}

\usepackage[all]{hypcap} 

\usepackage{xspace} 
\usepackage{upgreek}

\def\lhcb   {\mbox{LHCb}\xspace}

\def\MagUp {\mbox{\em Mag\kern -0.05em Up}\xspace}

\ifthenelse{\boolean{uprightparticles}}%
{

 \def\Ppi         {\ensuremath{\uppi}\xspace}

 \def\PDelta      {\ensuremath{\Delta}\xspace}                 
 \def\PXi         {\ensuremath{\Xi}\xspace}                 
 \def\PLambda     {\ensuremath{\Lambda}\xspace}                 
 \def\PSigma      {\ensuremath{\Sigma}\xspace}                 
 \def\POmega      {\ensuremath{\Omega}\xspace}                 
 \def\PUpsilon    {\ensuremath{\Upsilon}\xspace}
 \let\oldPi\Pi
 \def\PPi         {\ensuremath{\oldPi}\xspace}

 \def\PB      {\ensuremath{\mathrm{B}}\xspace}                 
                  
 \def\PD      {\ensuremath{\mathrm{D}}\xspace}

 \def\PK      {\ensuremath{\mathrm{K}}\xspace}

 \def\Pi      {\ensuremath{\mathrm{i}}\xspace}

 \def\Pp      {\ensuremath{\mathrm{p}}\xspace}

 \def\Ps      {\ensuremath{\mathrm{s}}\xspace}

 \def\thebaroffset{0.0em}
}
{

 \def\Ppi         {\ensuremath{\pi}\xspace}

 \mathchardef\PDelta="7101
 \mathchardef\PXi="7104
 \mathchardef\PLambda="7103
 \mathchardef\PSigma="7106
 \mathchardef\POmega="710A
 \mathchardef\PUpsilon="7107
 \mathchardef\PPi="7105
                  
 \def\PB      {\ensuremath{B}\xspace}                 
                  
 \def\PD      {\ensuremath{D}\xspace}

 \def\PK      {\ensuremath{K}\xspace}

 \def\Pi      {\ensuremath{i}\xspace}

 \def\Pp      {\ensuremath{p}\xspace}

 \def\Ps      {\ensuremath{s}\xspace}

 \def\thebaroffset{0.18em}
}
\newcommand{\offsetoverline}[2][\thebaroffset]{\kern #1\overline{\kern -#1 #2}}%

\makeatletter
\ifcase \@ptsize \relax
  \newcommand{\miniscule}{\@setfontsize\miniscule{4}{5}}
\or
  \newcommand{\miniscule}{\@setfontsize\miniscule{5}{6}}
\or
  \newcommand{\miniscule}{\@setfontsize\miniscule{5}{6}}
\fi
\makeatother

\DeclareRobustCommand{\optbar}[1]{\shortstack{{\miniscule (\rule[.5ex]{1.25em}{.18mm})}
  \\ [-.7ex] $#1$}}

\def\squark    {{\ensuremath{\Ps}}\xspace}

\def\pion   {{\ensuremath{\Ppi}}\xspace}

\def\KorKbar {\kern \thebaroffset\optbar{\kern -\thebaroffset \PK}{}\xspace}

\def\D       {{\ensuremath{\PD}}\xspace}

\def\DorDbar {\kern \thebaroffset\optbar{\kern -\thebaroffset \PD}\xspace}

\def\Dp      {{\ensuremath{\D^+}}\xspace}
\def\Dm      {{\ensuremath{\D^-}}\xspace}

\def\DpDm    {\ensuremath{\Dp {\kern -0.16em \Dm}}\xspace}

\def\B       {{\ensuremath{\PB}}\xspace}

\def\BorBbar {\kern \thebaroffset\optbar{\kern -\thebaroffset \PB}\xspace}

\def\Bd      {{\ensuremath{\B^0}}\xspace}

\def\BdorBdbar {\kern \thebaroffset\optbar{\kern -\thebaroffset \Bd}\xspace}

\def\Bs      {{\ensuremath{\B^0_\squark}}\xspace}

\def\BsorBsbar {\kern \thebaroffset\optbar{\kern -\thebaroffset \Bs}\xspace}

\def\Y#1S{\ensuremath{\PUpsilon{(#1S)}}\xspace}

\def\proton      {{\ensuremath{\Pp}}\xspace}

\def\LorLbar     {\kern \thebaroffset\optbar{\kern -\thebaroffset \PLambda}\xspace}

\def\AT#1     {\ensuremath{A_{\mathrm{T}}^{#1}}\xspace}           

\def\C#1      {\ensuremath{\mathcal{C}_{#1}}\xspace}                       
\def\Cp#1     {\ensuremath{\mathcal{C}_{#1}^{'}}\xspace}                    
\def\Ceff#1   {\ensuremath{\mathcal{C}_{#1}^{\mathrm{(eff)}}}\xspace}        
\def\Cpeff#1  {\ensuremath{\mathcal{C}_{#1}^{'\mathrm{(eff)}}}\xspace}       
\def\Ope#1    {\ensuremath{\mathcal{O}_{#1}}\xspace}                       
\def\Opep#1   {\ensuremath{\mathcal{O}_{#1}^{'}}\xspace}                    

\newcommand{\nospaceunit}[1]{\ensuremath{\text{#1}}}       
\newcommand{\aunit}[1]{\ensuremath{\text{\,#1}}}       

\newcommand{\tev}{\aunit{Te\kern -0.1em V}\xspace}
\newcommand{\gev}{\aunit{Ge\kern -0.1em V}\xspace}
\newcommand{\mev}{\aunit{Me\kern -0.1em V}\xspace}
\newcommand{\kev}{\aunit{ke\kern -0.1em V}\xspace}
\newcommand{\ev}{\aunit{e\kern -0.1em V}\xspace}
 
\newcommand{\mevc}{\ensuremath{\aunit{Me\kern -0.1em V\!/}c}\xspace}
\newcommand{\gevc}{\ensuremath{\aunit{Ge\kern -0.1em V\!/}c}\xspace}
\newcommand{\mevcc}{\ensuremath{\aunit{Me\kern -0.1em V\!/}c^2}\xspace}
\newcommand{\gevcc}{\ensuremath{\aunit{Ge\kern -0.1em V\!/}c^2}\xspace}

\def\mum  {\ensuremath{\,\upmu\nospaceunit{m}}\xspace}

\def\fb   {\ensuremath{\aunit{fb}}\xspace}
\def\invfb   {\ensuremath{\fb^{-1}}\xspace}

\def\deriv {\ensuremath{\mathrm{d}}}

\def\gsim{{~\raise.15em\hbox{$>$}\kern-.85em
          \lower.35em\hbox{$\sim$}~}\xspace}
\def\lsim{{~\raise.15em\hbox{$<$}\kern-.85em
          \lower.35em\hbox{$\sim$}~}\xspace}

\def\sqs   {\ensuremath{\protect\sqrt{s}}\xspace}
\def\sqsnn {\ensuremath{\protect\sqrt{s_{\scriptscriptstyle\text{NN}}}}\xspace}
\def\pt         {\ensuremath{p_{\mathrm{T}}}\xspace}

\def\ptot       {\ensuremath{p}\xspace}

\def\geant      {\mbox{\textsc{Geant4}}\xspace}

\def\photos     {\mbox{\textsc{Photos}}\xspace}

\def\pythia     {\mbox{\textsc{Pythia}}\xspace}

\def\tell1  {TELL1\xspace}
\def\ukl1   {UKL1\xspace}

\newcommand{\eg}{\mbox{\itshape e.g.}\xspace}
\newcommand{\ie}{\mbox{\itshape i.e.}\xspace}

\newcommand{\lhcborcid}[1]{\href{https://orcid.org/#1}{\hspace*{0.1em}\raisebox{-0.45ex}{\includegraphics[width=1em]{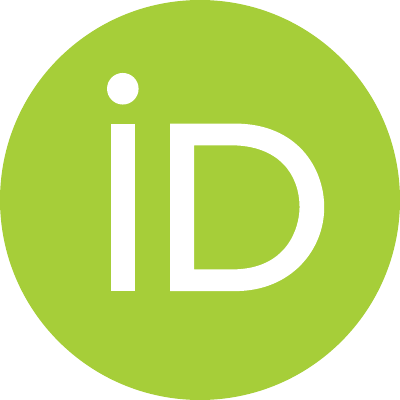}}}}

\hypersetup{
  colorlinks   = true, 
  urlcolor     = blue, 
  linkcolor    = blue, 
  citecolor    = red   
}

\ifEnableSectionTOCLinks
    \usepackage[explicit]{titlesec} 
    
    \let\oldcontentsline\contentsline
    \renewcommand

    \titleformat{\section}{\normalfont\Large\bf}{\hyperlink{tocsection.\thesection}{{\thesection} \parbox[t]{\dimexpr\textwidth-1pc}{#1}}}{1pc}{}

    \titleformat{\subsection}{\normalfont\bf}{\hyperlink{tocsubsection.\thesubsection}{{\thesubsection} \parbox[t]{\dimexpr\textwidth-1pc}{#1}}}{1pc}{}

    \titleformat{name=\section,numberless}[display]{}{}{0pt}{\normalfont\Huge\bfseries #1}
\fi

\usepackage{cite} 
\usepackage{mciteplus}

\usepackage{longtable} 

\newcommand{\pp}{\mbox{\proton\proton}\xspace}

\newcommand{\becQ}{\ensuremath{Q}\xspace}

\begin{document}

\renewcommand{\thefootnote}{\fnsymbol{footnote}}
\setcounter{footnote}{1}


\begin{titlepage}
\pagenumbering{roman}

\vspace*{-1.5cm}
\centerline{\large EUROPEAN ORGANIZATION FOR NUCLEAR RESEARCH (CERN)}
\vspace*{1.5cm}
\noindent
\begin{tabular*}{\linewidth}{lc@{\extracolsep{\fill}}r@{\extracolsep{0pt}}}
\ifthenelse{\boolean{pdflatex}}
{\vspace*{-1.5cm}\mbox{\!\!\!\includegraphics[width=.14\textwidth]{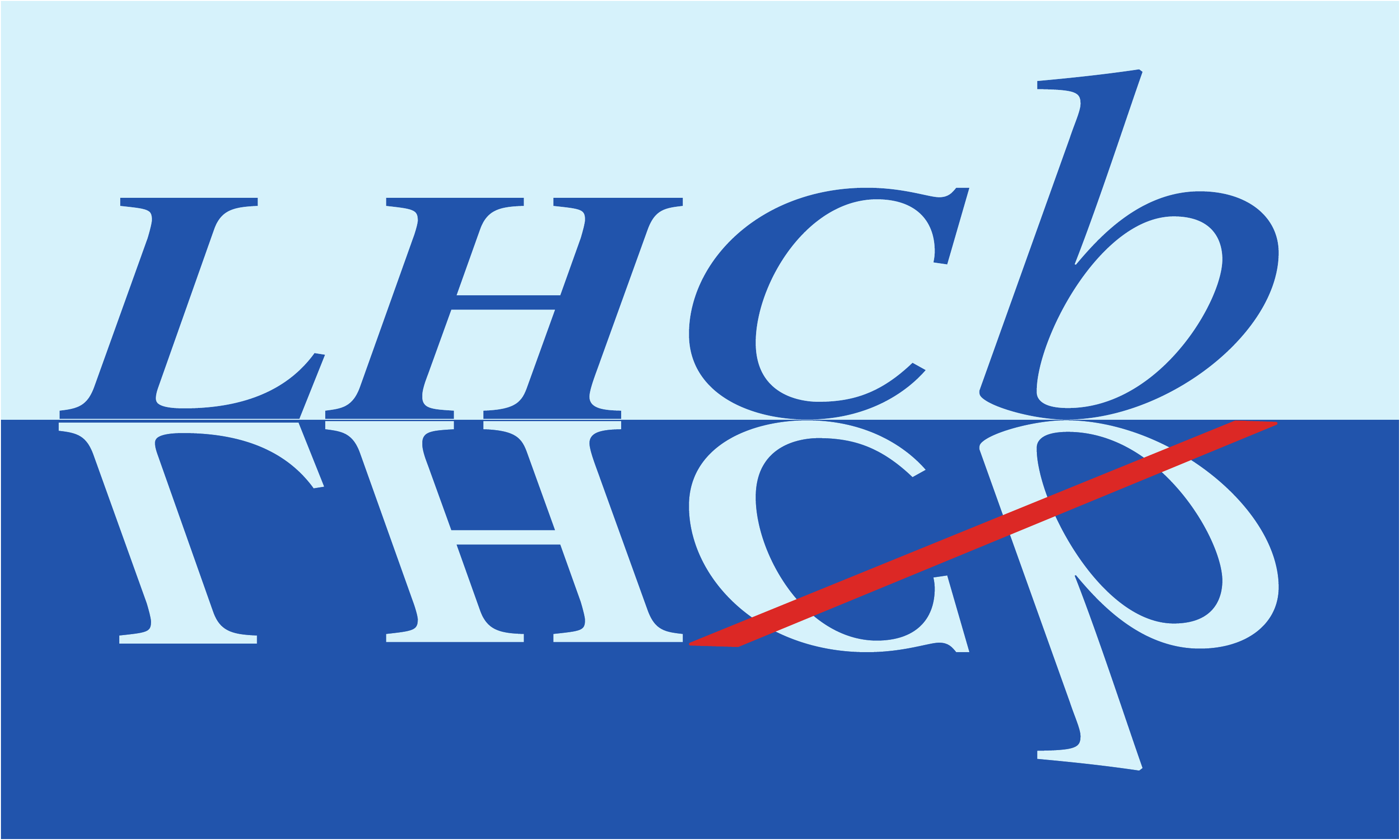}} & &}%
{\vspace*{-1.2cm}\mbox{\!\!\!\includegraphics[width=.12\textwidth]{figs/lhcb-logo.pdf}} & &}%
\\
 & & CERN-EP-2025-104 \\  
 & & LHCb-PAPER-2025-007 \\  
 & & August 28, 2025 \\ 
 & & \\
\end{tabular*}

\vspace*{4.0cm}

{\normalfont\bfseries\boldmath\huge
\begin{center}
  \papertitle 
\end{center}
}

\vspace*{2.0cm}

\begin{center}
\paperauthors
\footnote{Authors are listed at the end of this paper.}
\end{center}

\vspace{\fill}

\begin{abstract}
\noindent A study on the Bose--Einstein correlations for triplets of same-sign pions is presented. The analysis is performed using proton-proton collisions at a centre-of-mass energy of \sqs = 7\tev, recorded by the \lhcb experiment, corresponding to an integrated luminosity of 1.0\invfb. For the first time, the results are interpreted in the core-halo model. The parameters of the model are determined in regions of charged-particle multiplicity. This measurement provides insight into the nature of hadronisation in terms of coherence, being consistent with the presence of coherent emission of pions.
\end{abstract}

\vspace*{2.0cm}

\begin{center}
  Published in JHEP 08 (2025) 174
\end{center}

\vspace{\fill}

{\footnotesize 
\centerline{\copyright~\papercopyright. \href{\paperlicenceurl}{\paperlicence}.}}
\vspace*{2mm}

\end{titlepage}


\newpage
\setcounter{page}{2}
\mbox{~}
%
%
%
%


\renewcommand{\thefootnote}{\arabic{footnote}}
\setcounter{footnote}{0}

\cleardoublepage


\pagestyle{plain} 
\setcounter{page}{1}
\pagenumbering{arabic}


\section{Introduction}
\label{sec:Introduction}

In particle and nuclear physics, intensity interferometry offers a direct experimental approach to determine the size, shape, and lifetime of particle-emitting sources~\cite{HanburyBrown:1954,Brown:1956,HanburyBrown:1956}. Specifically, boson interferometry is a valuable tool for studying the space-time structure of particle production processes. This is because Bose--Einstein correlations (BEC)~\cite{Brown:1956} between two or three identical bosons reveal both the geometric and dynamic characteristics of the particle-emitting source. The correlations result from quantum statistics and the final-state interactions of strong and electromagnetic origins. When identical pions are produced close together in phase space, their quantum wavefunctions interfere constructively~\cite{HanburyBrown:1954}, enhancing the probability of detecting pion pairs with small relative momenta. The space-time properties of the hadron-emission volume can be studied through the parameters of the density function in the region of small four-momenta difference using the quantum interference effect between indistinguishable particles emitted by a finite-size source. Such measurements are also often referred to as femtoscopy~\cite{Lisa_2005}. The so-called small systems, \ie proton-proton (\pp) and proton-ion collisions, are characterised by significantly shorter lifetimes than their ion-ion counterparts~\cite{Schenke:2014}, giving a better experimental insight into the early system dynamics and the initial geometry~\cite{Schenke:2014,s10052-015-3509-3}. The available accuracy and data-sample sizes in various high-energy physics experiments make it possible to study three-particle correlations, which may reveal the nature of the hadronisation stage and support the validation of theoretical models~\cite{Bialas:2014gca, GOTTSCHALK1984325, Csorgo:1994in}. The main interest in three-particle correlations is the study of hadron creation mechanisms beyond chaotic emission as opposed to two-particle correlations, sensitive to thermalisation. Study of two- and three-particle correlations can be used to experimentally measure thermalisation and coherence in the source.

Previous results on three-particle correlations have already been reported by the ALICE experiment in PbPb collisions at the LHC~\cite{PhysRevC.89.024911}. However, with a limited interpretation of the results within the core-halo model~\cite{Csorgo:1994in}. Recently, measurements of three-particle correlations in AuAu collisions were interpreted within the core-halo model (see Sec.~\ref{sec:Core-halo}) at the PHENIX experiment~\cite{Phenix3BEC,art13}. This model provides a set of parameters describing the expansion of the system after a collision and describes hadronic source properties not available in two-particle correlation analyses.

This paper reports for the first time the measurements of the three-particle correlations in small collision systems, performed in different regions of charged-particle multiplicity, employing \pp collisions at \sqs = 7\tev recorded by the \lhcb experiment, corresponding to an integrated luminosity of 1\,fb$^{-1}$. A convolution of the information from the two-particle correlation function, determined in a previous LHCb analysis~\cite{LHCb-PAPER-2017-025}, with the  three-particle BEC, may probe the parameters of the core-halo model~\cite{Csorgo:1994in}, disentangling the partial coherence and the halo contribution. In particular, the parameters can be studied in relation to a thermalised core. To achieve this, a new parameter is introduced, $\kappa_3$~\cite{Phenix3BEC}, being a function of both two- and three-particle correlation strengths, $\lambda_2$ and $\lambda_3$~\cite{Phenix3BEC,Csorgo2000}. This new parameter is different from unity when additional effects are present within the core, such as an imperfectly thermalised core or partial coherence. The analysis investigates whether $\kappa_3$ reflects such additional effects. The values of the partial coherence parameter $p_c$ and the fraction of the particles that originate from the core $f_c$ are also measured as a function of the charged-particle multiplicity. A key question is whether coherent emission can also be observed in small collision systems like \pp, which would manifest in the multiplicity-dependence of $p_c$ and $\kappa_3$ parameters.

The paper is organised as follows. In Sec.~\ref{sec:Analysis} an overview of the theory regarding the two- and three-particle correlations is given, followed by an introduction of the core-halo model used in the present analysis. The LHCb detector and data samples used are described in Sec.~\ref{Detector}, while the event selection and the fits to the three-pion correlation functions are reported in Sec.~\ref{sec:evtSel} and~\ref{finalFits}, respectively. The systematic uncertainties and final results are presented in Secs.~\ref{sec:systematics} and~\ref{results}. A summary of the studies performed is given in Sec.~\ref{conclusions}.

\section{Theoretical framework}
\label{sec:Analysis}

The BEC effect arises from quantum statistics, resulting from the symmetrisation of the wavefunction describing a system of bosons. The present study is based on the assumption of a static, spherically symmetric source that can be characterised by univariate distributions. To measure the three-particle correlation strength $\lambda_3$ as well as the parameters of the core-halo model, the three-particle correlation function is measured, which is a convolution of two-particle correlation functions for each particle pair within a given triplet. In the present analysis the correlation radius $R$ and the two-particle correlation strength $\lambda_2$ are obtained from the two-particle correlation studies~\cite{LHCb-PAPER-2017-025}. For consistency with the published two-pion BEC analysis, the same event data set is used, collected in 2011 at a centre-of-mass energy of \sqs = 7\tev.

\subsection{Two-particle correlation function}
\label{sec:CF2}

The correlation function is commonly studied using the Lorentz-invariant variable $Q$~\cite{Baym:1997}, which is related to the difference in the four-momenta $k_{1}$ and $k_{2}$ of two indistinguishable particles of rest mass $m$ by
\begin{equation}
	\label{eq:theory:defQ}
	\becQ \equiv \sqrt{ - \left(k_{1} - k_{2} \right)^2} = \sqrt{M^{2} - 4 m^{2}}\ .
\end{equation}
This provides a measure of the phase-space separation within the two-particle system of invariant mass $M$. The two-particle correlation function $C_{2}$ is formulated as the ratio of the $Q$ distributions for signal and reference pairs,
\begin{equation}
    C_2(Q) \equiv \frac{\rho_2(Q)^{\text{sig}}}{\rho_2(Q)^{\text{ref}}} = \left( \frac{N^{\text{ref}}}{N^{\text{sig}}} \right) \left( \frac{ \deriv N^{\text{sig}}(Q) \, / \, \deriv Q }{ \deriv N^{\text{ref}}(Q) \, / \, \deriv Q } \right) \ ,
\end{equation}
where $\rho_2(Q)^{\text{sig}}$ represents the $Q$ distribution of the same-charge (same-sign) particle pairs, while $\rho_2(Q)^{\text{ref}}$ represents the reference sample without the BEC effect. Here $N^{\text{sig}}$ and $N^{\text{ref}}$ correspond to the~number of signal and reference pairs, respectively, obtained from an~integral of the~relevant $Q$ distributions. In this study, the reference sample without the BEC effect is created by pairing particles from different events in data using the event-mixing method~\cite{Alexander:2003ug}. Reference pairs are selected similarly to signal pairs to ensure that the kinematic distributions agree. Specifically, the particles in the reference pairs must come from primary vertices with a comparable number of tracks reconstructed in the vertex detector (see Sec.~\ref{Detector}). 

In the case of static, univariate sources, the two-particle correlation function is commonly parametrised with the L\'evy-type function~\cite{BECforLevyStable}
\begin{equation}
\label{eq:cff}
    C_2(Q) = N(1 + \lambda_2 e^{-(RQ)^{\alpha_{\text{\miniscule L}}}}) (1 + \delta Q).
\end{equation}
The value of $R$ can be interpreted as the radius of the spherically symmetric source of the pion-emission volume, and $N$ is the overall normalisation factor. The intercept parameter $\lambda_2$ corresponds to the extrapolated value of the correlation function at $Q = 0$\gev~\cite{Csorgo2000}.\footnote{If not indicated otherwise, natural units are used throughout the paper.} This parameter is also referred to as the correlation strength. The parameter $\delta$ corresponds to the long-range momentum correlations. The value of the L\'evy index of stability $\alpha_{\text{\tiny L}}$ depends on the assumed density distribution~\cite{BECforLevyStable} and it can take values in the range $(0,2]$. 

The effects related, among others, to the imperfections of the construction of the reference sample, can be mitigated  using the so-called \emph{double ratio}, a technique commonly used in BEC studies~\cite{Abbiendi:2000jb}. The ratio of the correlation functions for data and simulation, used in the final fit, is expressed as
\begin{equation}
    \label{eq:doubleratio}
    r_{\rm d}(Q) \equiv \frac{C_2^{\text{data}}(Q)}{C_2^{\text{sim}}(Q)}.
\end{equation}
The correlation function is constructed for the simulated sample with the BEC effect turned off. The double ratio is  used to suppress some nonfemtoscopic effects. It applies only to well-simulated phenomena, such as long-range correlations, and it is important to maintain the same set of selection parameters for both samples.

\subsection{Three-particle correlation function}
\label{sec:CF}

The three-particle correlation function is defined as the ratio of the three-particle four-momentum distribution to the product of one-particle distributions
\begin{equation}
    C_3(k_1,k_2,k_3) \equiv \frac{N_3(k_1,k_2,k_3)}{N(k_1)N(k_2)N(k_3)},
\end{equation}
where the single-particle four-momentum distribution is defined as $N(k) = \int_{}^{} \deriv^4x S(x,k)$ and $S(x,k)$ is the source distribution, which describes the probability density of particle creation at the spacetime point $x$ with four-momentum $k$. The three-particle distribution is defined as
\begin{equation}
    N_3(k_1,k_2,k_3) \equiv \int_{}^{} \deriv^4 x_1 \deriv^4 x_2 \deriv^4 x_3 S(x_1, k_1) S(x_2, k_2) S(x_3, k_3) |\Psi_{k_1,k_2,k_3} (x_1,x_2,x_3)|^2,
\end{equation}
with $\Psi_{k_1,k_2,k_3}$ being the three-particle wavefunction.

Assuming properly symmetrised plane-waves for the wavefunctions and the L\'evy distribution as the source function, the three-particle correlation function can be expressed as a convolution of two-particle correlation functions for each pair of particles from the triplet~\cite{Phenix3BEC,BECforLevyStable}
\begin{equation}
    \label{eq:C3_corehalofit}
    \begin{split}
    C_3^{(0)}(Q_{12},Q_{13},Q_{23}) = C_{2}^{(0)}(Q_{12}) C_{2}^{(0)}(Q_{13}) C_{2}^{(0)}(Q_{23}) = \qquad \qquad \qquad \quad \quad \quad \qquad \\ 1 + \ell_3 e^{-0.5[(Q_{12}R)^{\alpha_{\text{\miniscule L}}} + (Q_{13}R)^{\alpha_{\text{\miniscule L}}} + (Q_{23}R)^{\alpha_{\text{\miniscule L}}}]}+ \ell_2[e^{-(Q_{12}R)^{\alpha_{\text{\miniscule L}}}} + e^{-(Q_{13}R)^{\alpha_{\text{\miniscule L}}}} + e^{-(Q_{23}R)^{\alpha_{\text{\miniscule L}}}} ],
    \end{split}
\end{equation}
where $Q_{mn} = \sqrt{-(k_m - k_n)^2}$ is the four-momentum difference, and the subscript labels the particle in the triplet. The upper script (0) describes correlation functions with the pure BEC effect. The L\'evy index of stability $\alpha_{\text{\tiny L}}$ is set to $1$ to be compatible with the two-particle correlation analysis results~\cite{LHCb-PAPER-2017-025} from which the $R$ and $\lambda_2$ parameters are taken. The three-particle correlation strength is defined as $\lambda_3 \equiv \ell_3 + 3 \ell_2$~\cite{Phenix3BEC}. The fitted three-pion double ratio is defined as $r_{\rm d_{3}}(Q) \equiv C_3^{\text{data}}(Q) / C_3^{\text{sim}}(Q)$.

\subsection{Final-state interactions and nonfemtoscopic background}
\label{sec:fsi}

Additional effects, known as final-state interactions, can affect the correlation function. These effects are associated with strong and electromagnetic forces. The effect of the strong interactions in the case of pions is relatively small~\cite{wil2, wil3} and is usually neglected in BEC studies. The more significant is Coulomb repulsion related to the same-sign electric charge of particles being investigated, affecting the correlations, especially in the low-$Q$ region. For point-like sources, the Coulomb interaction is equivalent to the so-called Gamov penetration factor~\cite{PhysRevC.20.2267, riverside, PhysRevD.33.72},
\begin{equation}
    \label{eq:gamow}
    G_2(Q) = \frac{2 \pi \zeta}{e^{2 \pi \zeta} -1}, \quad \zeta = \pm \frac{\alpha m}{Q},
\end{equation}
where $\alpha$ is the fine-structure constant and the sign of $\zeta$ is positive for the same-sign particles, and negative otherwise.

Long-range correlations, being one of the nonfemtoscopic effects~\cite{Sirunyan:2017} related mainly to energy-momentum conservation, are largely reduced by using the double-ratio method and are accounted for through the introduction of the $\delta$ parameter with $N$ serving as the normalisation 
\begin{equation}
    \begin{split}
        C_3(Q_{12}, Q_{13}, Q_{23}) &= \\ N(1+\delta_{12} Q_{12})(1+&\delta_{13} Q_{13})(1+\delta_{23} Q_{23}) G_3(Q_{12}, Q_{13}, Q_{23}) C_3^{(0)}(Q_{12}, Q_{13}, Q_{23}),  
    \end{split}
\end{equation}   
where $G_3$ represents the correction of the Coulomb effect. According to the generalised Riverside method~\cite{riverside,riverside2}, the Coulomb correction for the particle triplet can be approximated by a factorization of the corrections calculated for each of the pairs in the triplet $G_3(Q_{12}, Q_{13}, Q_{23}) \approx G_2(Q_{12}) G_2(Q_{13}) G_2(Q_{23})$. It is important to note that the Coulomb effect is not present in the simulated data, hence, the correction is applied only to the data.

The effect referred to as the cluster contribution~\cite{Sirunyan:2017} is another prominent issue and is related to minijet fragmentations and multibody resonance decays. It is more challenging to correct than long-range correlations as it appears in the range of $Q < 1.0$\gev that overlaps with the BEC signal. However, this contribution is not expected to be prominent in proton-proton collisions, as the double-ratio distributions for opposite-sign pion pairs are almost uniform and close to unity~\cite{LHCb-PAPER-2017-025}. This indicates that the nonfemtoscopic background is properly modelled in simulation and corrected for when the double ratio is used. Potential discrepancies related to the event generator used to simulate the same-sign and opposite-sign pairs are addressed in the systematic uncertainty studies.

\subsection{Core-halo model}
\label{sec:Core-halo}
The region of pion emission after a proton-proton collision can be described as a system with a large halo \cite{Csorgo:1994in}. Depending on the characteristics of the hadron emission, the volume can be divided into two distinguishable parts -- the core and the halo. In the central part, the \emph{core}, particles are created in direct processes of the hydrodynamic evolution or particle production from excited strings and following rescattering. The \emph{halo} around the core consists of pions originating from decays of long-lived hadronic resonances, \eg $\eta$, $\eta'$, $\omega$ and $K^0$, where long-lived means a decay length over 20 fm~\cite{Csorgo:1994in}. In the case of a relatively small core, the dimensions of the core can be determined using Bose--Einstein correlations. As the model accounts also for the partial coherence in the core, both the partial coherence and the halo contribution constrain the value of the two-particle Bose--Einstein correlation function to below 2 at $Q = 0$\gev. Therefore, in order to disentangle the partial coherence and the halo contribution, two separate measurements are necessary, the two- and three-particle Bose--Einstein correlation. The parameters of the core-halo system can be used to describe the properties of the hadron emission. 

The fraction of the core, $f_c$, which quantifies the portion of particles that originated from the core, together with the partial coherence parameter, $p_c$, which represents the fraction of particles emitted coherently, can be used to express the correlation strengths~\cite{Csorgo2000}
\begin{equation}
 \label{eq:corehaloLambda2}
    f_c^2[(1-p_c)^2 + 2p_c(1-p_c)] = \lambda_2,
\end{equation}
\begin{equation}
 \label{eq:corehaloLambda3}
   2f_c^3[(1-p_c)^3 + 3p_c(1-p_c)^2] + 3f_c^2[(1-p_c)^2 + 2p_c(1-p_c)] = \lambda_3,
\end{equation}
where $\lambda_2$ is taken from the published two-pion BEC analysis~\cite{LHCb-PAPER-2017-025}, while $\lambda_3$ is determined based on fitted parameters (see Sec.~\ref{sec:CF}).

Experimentally, to be able to investigate the partial coherence within the limits of the core-halo model, Eqs.~\ref{eq:corehaloLambda2} and \ref{eq:corehaloLambda3} should be solved for $f_c$ and $p_c$ using the measured values of the $\lambda_2$ and $\lambda_3$ parameters as input.

The measurement of $\lambda_2$, in conjunction with $\lambda_3$, has the potential to challenge the boundaries of the core-halo model~\cite{Csorgo:1994in} with a thermalised core. Consequently, a novel observable, $\kappa_3$, is introduced~\cite{Phenix3BEC}, being a function of $\lambda_3$ and $\lambda_2$,
\begin{equation}    \label{eq:corehaloKappa3}
    \kappa_3 = 0.5(\lambda_3 - 3\lambda_2)/\lambda_2^{3/2}.
\end{equation}
This parameter is independent of the fraction of particles originating from the core, and can be used to determine additional effects in the core, such as partial coherence or a not fully thermalised core. Its value is unity if the particle emission has no coherent component~\cite{BECforLevyStable,art13}.

\subsection{Parameterisation of the three-particle correlation function}
\label{sec:Fitting}

The experimentally measurable three-particle correlation function, corrected for the Coulomb effect and including long-range correlations, is modelled by

\begin{equation}
    \label{eq:3BECfullparam}
    \begin{split}
        C_3(Q_{12}, Q_{13}, Q_{23}) = N(1+\delta_{12} Q_{12})(1+\delta_{13} Q_{13})(1+\delta_{23} Q_{23}) \\ (1 + \ell_3 e^{-0.5 R(Q_{12} + Q_{13} + Q_{23})} + \ell_2(e^{-Q_{12}R} + e^{-Q_{13}R} + e^{-Q_{23}R})),
    \end{split}
\end{equation}
where the L\'evy index of stability $\alpha_{\text{\tiny L}}$ is fixed to 1, and where $\ell_2$ and $\ell_3$ are parameters related to the two-particle correlation strength $\lambda_2$ and three-particle correlation strength $\lambda_3$~\cite{Phenix3BEC}, giving information about the hadron creation mechanism. 

Events are selected with the four-momenta differences satisfying $Q_{12} \approx Q_{13} \approx Q_{23}$ to determine $\lambda_3$ and the values of the parameters of the core-halo model at a specific value of $Q$. Both parameters are related to the correlation strengths in the way described in Eqs.~\ref{eq:corehaloLambda2} and \ref{eq:corehaloLambda3}. The third parameter of this model, $\kappa_3$, can also be expressed using correlation strengths, as demonstrated in Eq.~\ref{eq:corehaloKappa3}. The correlation radius $R$ as well as the $\lambda_2$ parameters are obtained from the previous two-particle correlation analysis~\cite{LHCb-PAPER-2017-025}.

\section{Detector and dataset}
\label{Detector}

The LHCb detector~\cite{LHCb-DP-2008-001,LHCb-DP-2014-002} is a single-arm forward spectrometer designed to study particles containing $b$ or $c$ quarks. It features a high-precision tracking system, which includes a silicon-strip vertex detector (VELO)~\cite{LHCb-DP-2014-001} surrounding the \pp interaction region, covering the pseudorapidity range $2 < \eta < 5$.\footnote{Pseudorapidity is the variable that corresponds to the polar angle $\theta$ of the track with respect to the beam axis, $\eta = -\ln(\tan{(\theta/2))}$. References to pseudorapidity in the present study correspond to the values in the LHCb laboratory frame. The LHCb coordinate system is right-handed, with the $z$-axis pointing along the beam axis, $y$-axis in the vertical direction, and $x$-axis in the horizontal direction. The ($x,z$)-plane is the bending plane of the dipole magnet.} Upstream of a dipole magnet with a bending power of about 4\,T\,m, a large-area silicon-strip detector is located, along with three stations of silicon-strip detectors and straw drift tubes~\cite{LHCb-DP-2013-003} downstream of the magnet. The tracking system measures the momentum, \ptot, with a relative uncertainty ranging from 0.5\% at low momentum to 1\% at 200\gev. The impact parameter, or the minimum distance of a track to a primary vertex (PV), is measured with a resolution of (15 + 29/\pt)\mum, where \pt is the transverse component of \ptot in\gev. Charged hadrons are identified using information from two ring-imaging Cherenkov detectors~\cite{LHCb-DP-2012-003}. Photons, electrons and hadrons are identified by a calorimeter system. Muons are detected by a system of alternating layers of iron and multiwire proportional chambers~\cite{LHCb-DP-2012-002}. The trigger~\cite{LHCb-DP-2012-004} consists of a hardware stage based on information from the calorimeter and muon systems, followed by a software stage that performs full event reconstruction.

In this analysis, a \pp collision dataset of no-bias and minimum-bias triggered events is used, collected in 2011 at a centre-of-mass energy of \sqs = 7\tev, corresponding to an integrated luminosity of 1.0\invfb. The data sample available in the present study corresponds to $\sim$$4 \times 10^7$ events. The no-bias trigger selects events randomly, while the minimum-bias trigger requires at least one reconstructed VELO track.\footnote{A VELO track is reconstructed only with hits registered in the VELO detector.} The data were collected with an average number of visible interactions per bunch crossing (pile-up) of 1.4~\cite{LHCb-PAPER-2014-047}. To avoid biases related to trigger requirements, a sample of uncorrelated $pp$ interactions is constructed as described in Sec.~\ref{Detector}.

In the simulation, $pp$ collisions are generated using \pythia 8 \cite{Sjostrand:2007gs} with a specific LHCb configuration~\cite{LHCb-PROC-2010-056}, excluding the BEC effect. The decays of hadronic particles are simulated using EvtGen \cite{Lange:2001uf}, with final-state radiation generated by \photos~\cite{davidson2015photos}. The interaction of the generated particles with the detector and its response are modelled using the \geant toolkit~\cite{Agostinelli:2002hh, Allison:2006ve}, as described in~\cite{LHCb-PROC-2011-006}.\footnote{The version numbers of key simulation tools: PYTHIA 8.1, EvtGen 2.2, PHOTOS 2.0, Geant4 v70r0p1.} The simulated dataset contains ~$\sim$$2 \times 10^7$ minimum bias events. For the study of systematic effects, an additional sample of $\sim$$10^7$ minimum bias events is simulated using \pythia~6.4 \cite{Sjostrand:2006za} with the Perugia0 tune~\cite{Skands:2009zm}. 

In order to be compatible with the two-pion correlation analysis results~\cite{LHCb-PAPER-2017-025}, the same data and simulation samples, together with the same packages versions, are used.

\section{Event selection}
\label{sec:evtSel}

The analysis utilizes a sample of events that may include multiple $pp$ collisions. Without trigger requirements, each $pp$ interaction within an event can be analysed independently. Consequently, when an event is selected by the no-bias trigger, all primary vertices are included. For events with multiple $pp$ collisions selected by the minimum-bias trigger, any related biases are mitigated by randomly removing one of the PVs associated with the track(s) that triggered the event.

As the correlation parameters are expected to be dependent upon the particle multiplicity, the BEC effect is studied across different bins of reconstructed charged-particle multiplicity. Three \emph{activity classes} are introduced to reflect  the total multiplicity in the full solid angle. These classes are defined based on the VELO track multiplicity ($N^{\text{VELO}}_{\text{ch}}$) of the reconstructed primary vertices, which is a reliable analogue of the total multiplicity. The low-activity class consists of PVs with $N^{\text{VELO}}_{\text{ch}} \in [5,10]$ (48\% of all reconstructed PVs), the medium activity with $N^{\text{VELO}}_{\text{ch}} \in [11,20]$ (37\%) and the high activity with $N^{\text{VELO}}_{\text{ch}} \in [21,60]$ (15\%).

The following single-particle selection criteria are applied. All pion candidates must have reconstructed track segments in the VELO, with $2 < \eta < 5$, and in tracking stations downstream of the magnet. Each track is required to have a good-quality fit~\cite{LHCb-TDR-009}, a transverse momentum \pt $> 0.1$\gev, and no associated signal in the muon stations. Both pion candidates must be assigned to the same PV. Particles are assigned to the PV that minimises the $\chi^2$ value of the impact parameter, defined as the difference in the vertex-fit $\chi^2$ values of a given PV reconstructed with and without the track in question. A loose requirement on the track impact parameter ($<$ 0.4 mm) is applied to retain the majority of the particles originating from the selected PV. To reduce contamination from fake and clone tracks,\footnote{Fake tracks are those which do not correspond to any particle trajectory, but are reconstructed from a number of unrelated hits. Clone tracks are multiple tracks reconstructed from hits which were deposited by a single charged particle.} only the track with the best $\chi^2$ is retained when multiple tracks share the same hits in the VELO subdetector. Additionally, fake tracks are removed by applying requirements on the track $\chi^2$ and using the output of a dedicated neural network~\cite{LHCb-DP-2012-003}.

To mitigate the effect from clone tracks, the slopes of the tracks are studied, as clone tracks typically share very similar trajectories, resulting in small differences in the relevant slopes, defined as ratios of appropriate momentum components ($\Delta t_x = p_{x1} / p_{z1} - p_{x2} / p_{z2}$ and $\Delta t_y = p_{y1} / p_{z1} - p_{y2} / p_{z2}$). If both $|\Delta t_x|$ and $|\Delta t_y|$ are less than $0.3 \times 10^{-3}$ , the pair is discarded. After applying such requirements, the effect of clone particles is found to be negligible in the region $Q > 0.05$\gev.

Particle identification (PID) is performed using a neural network that incorporates subdetector information to calculate the probability for a particle to be identified as a specific type~\cite{LHCb-DP-2012-003}. The simulated PID quantities are corrected using calibration samples from the data~\cite{LHCb-PUB-2016-021}. Corrections of PID variables among particles are also taken into account. It is crucial to maintain a sample with high purity; however, applying a strict requirement on the probability for a particle to be identified as a pion, $\cal{P}(\pion)$, could suppress low-momentum pions that contribute to the BEC effect and significantly impact the signal region of the correlation function. To ensure consistency with previous two-pion analyses of \pp collisions~\cite{LHCb-PAPER-2017-025}, the requirement of $\cal{P}(\pion)$ $>$ 0.65 is applied. With this relatively loose selection criterion, the single-track pion identification purity is approximately 98\%. This leads to a pion pair purity of 96\%. The contamination from mixed pairs containing only one true pion is about 4\%. This level of contamination has been verified to have no impact on the correlation function, as the Bose--Einstein correlation effect does not occur in such cases. The total contribution from same-type particle pairs not containing pions is below 10$^{-4}$. Varying $\cal{P}(\pion)$ threshold from 0.5 to 0.8 does not result in significant changes in the measured correlation function.

\section{Fitting the correlation function}
\label{finalFits}

Correlation functions are constructed for $Q$ values in the range 0.05--2.00\gev with a bin width of 0.005\gev for consistency with the two-pion BEC analysis~\cite{LHCb-PAPER-2017-025}. In order to collect unique triplets of the same-sign charged pions coming from the same PV, three same-sign pions are randomly selected from the initial set of such particles. Used pions are then removed from the set. The reference sample is prepared using the event-mixing method. To ensure that the correlations are absent in reference triplets, each particle is chosen from a different event, from a PV with the same charged particle multiplicity. The reference triplets are processed in the same way as the signal ones, to ensure that the signal distribution is reproduced as closely as possible. 

As the final results for the three-particle correlation measurement depend on the $\lambda_2$ and $R$ parameters as measured with two-particle correlations, the parameters for the pairs from particle triplets were checked to be compatible with the previous LHCb results for two-particle BEC~\cite{LHCb-PAPER-2017-025}.

The calculation of the three-pion correlation function is based on the convolution of the two-pion correlation functions, where the pion pairs must originate from the triplet of the same-sign pions assigned to the same PV. First, the double ratios of the two-particle correlation functions (as outlined in~Eq.~\ref{eq:doubleratio}) are fitted separately for the three Coulomb-corrected two-particle double ratios for $Q_{12}$, $Q_{23}$ and $Q_{13}$ using the formula for the two-particle correlations (Eq.~\ref{eq:cff}) with $\alpha_{\text{\tiny L}} = 1$, 
\begin{equation}
    \label{eq:2BEC}
    C_2(Q) = N(1 + \lambda_2 e^{-RQ}) (1 + \delta Q).
\end{equation}
The obtained results for the $R$ and $\lambda_2$ parameters from these fits are then compared to the values published in Ref.~\cite{LHCb-PAPER-2017-025} in order to verify their compatibility. The values of $R$ and $\lambda_2$ are found to be consistent in all bins of VELO track multiplicity.

Finally, the distributions of the double ratio of correlation functions for same-sign pion triplets, $r_{\rm d_{3}}$, using the event-mixed reference sample (as described in Sec.~\ref{sec:CF2}), are fitted. The fit is performed in three different regions (bins) of VELO track multiplicity per \pp collision using the parameterisation in Eq.~\ref{eq:3BECfullparam} with four-momenta differences $Q_{12}$, $Q_{13}$, $Q_{23}$ selected to fall into the same $Q$-bin. A binned maximum-likelihood fit is used. The fitted parameters are $N$, $\delta$, $\ell_2$ and $\ell_3$, while the values of $R$ and $\lambda_{2}$ are taken from the published two-pion BEC analysis~\cite{LHCb-PAPER-2017-025}, whose values are, depending of $N^{\text{VELO}}_{\text{ch}}$ bin, in the range 1.01--1.80~fm and 0.57--0.72, respectively. The results of the fits to the three-particle double ratios are presented in Fig.~\ref{fig:3BEC_fits}, while the related $\chi^{2}/ndf$ can be found in Appendix~\ref{corr}.  The Bose--Einstein correlation effect is evident as a signal enhancement for low values of $Q$. It is worth noting that the fit quality in BEC studies is not expected to be perfect. Due to the lack of theoretical description of nonfemtoscopic background contributions as well as the compromise between the fit quality and interpretability of the measured correlation parameters, the obtained values of $\chi^2$ per degree of freedom are often larger than unity.

\begin{figure}
    \begin{center}
        \includegraphics[width=0.49\textwidth]{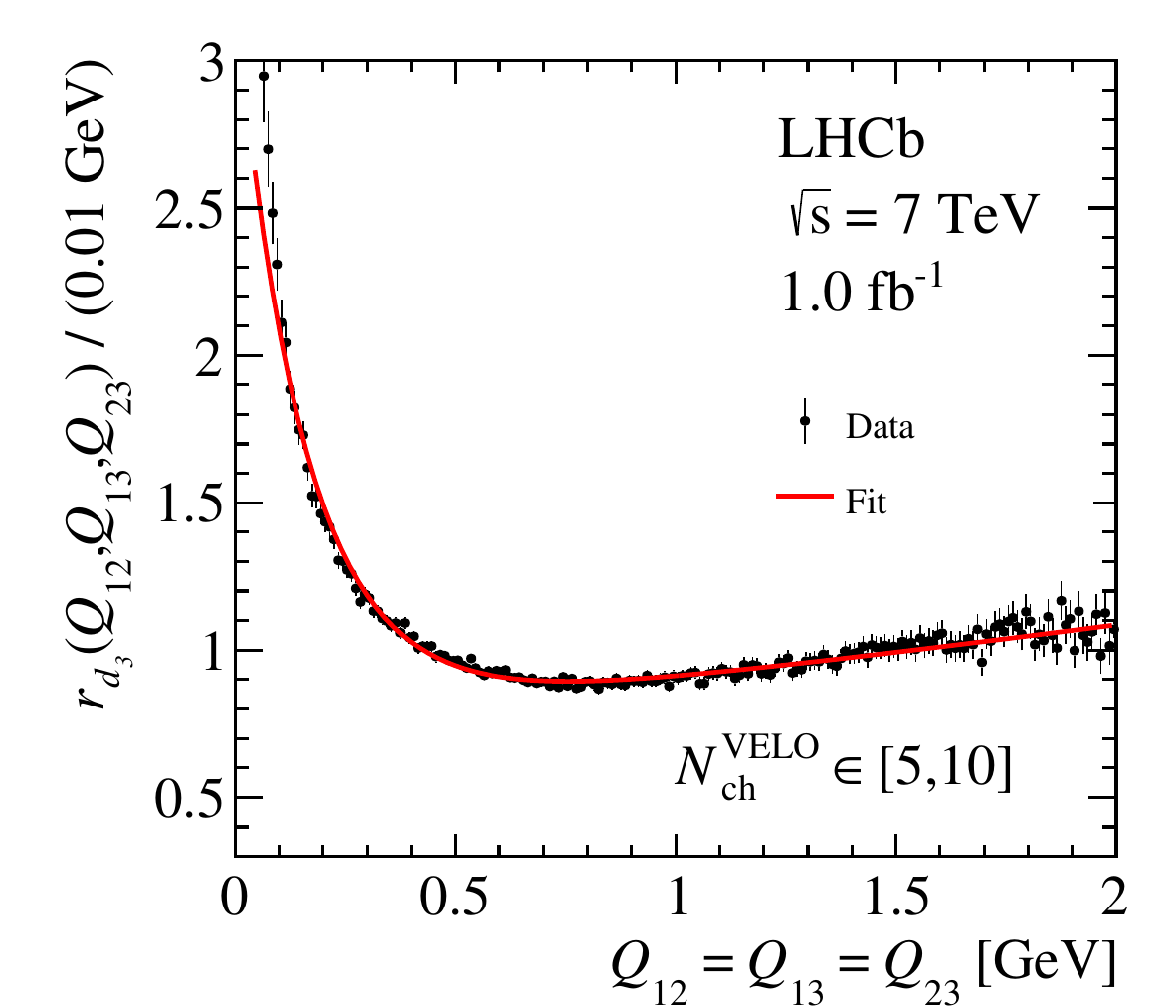} \hfill
        \includegraphics[width=0.49\textwidth]{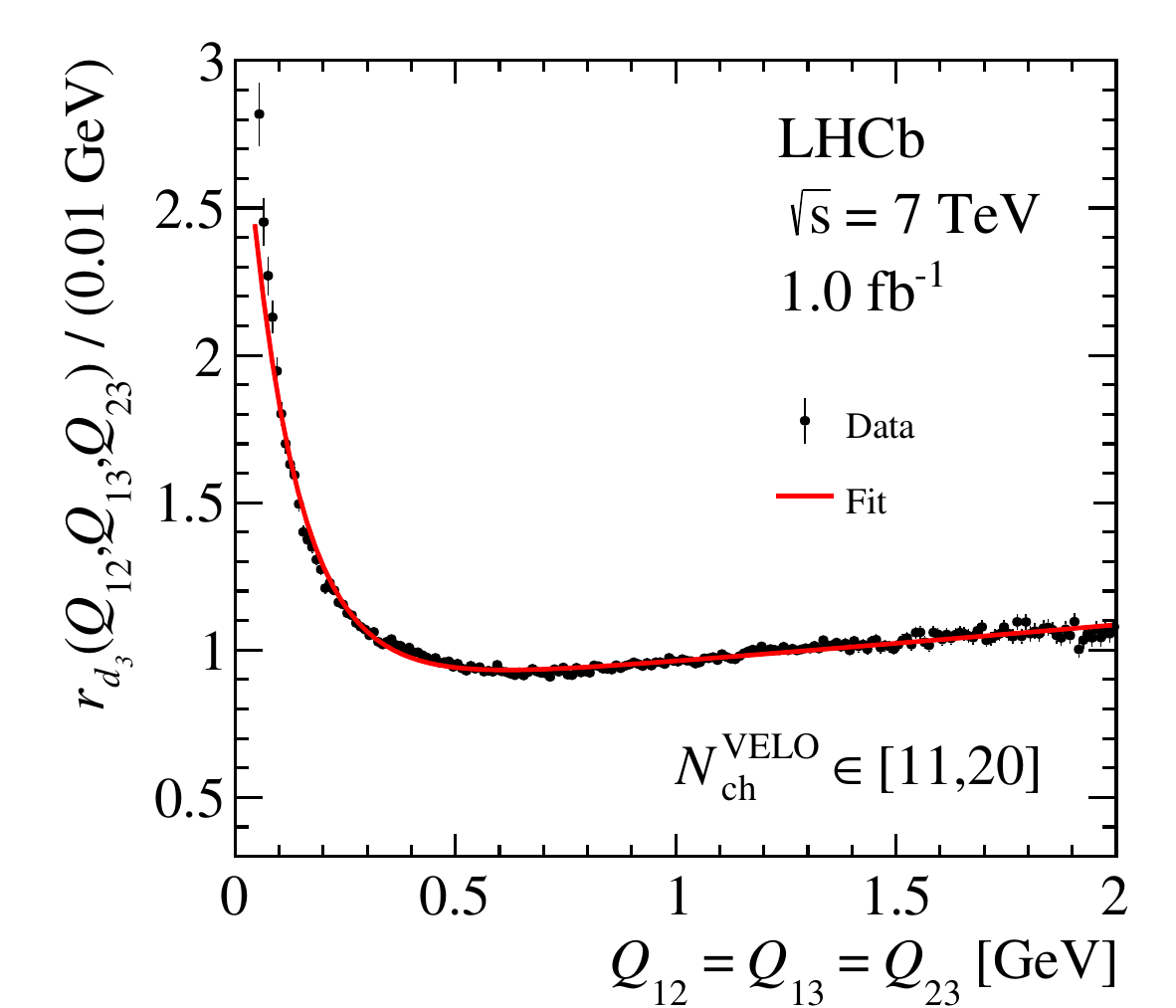} \\
        \includegraphics[width=0.49\textwidth]{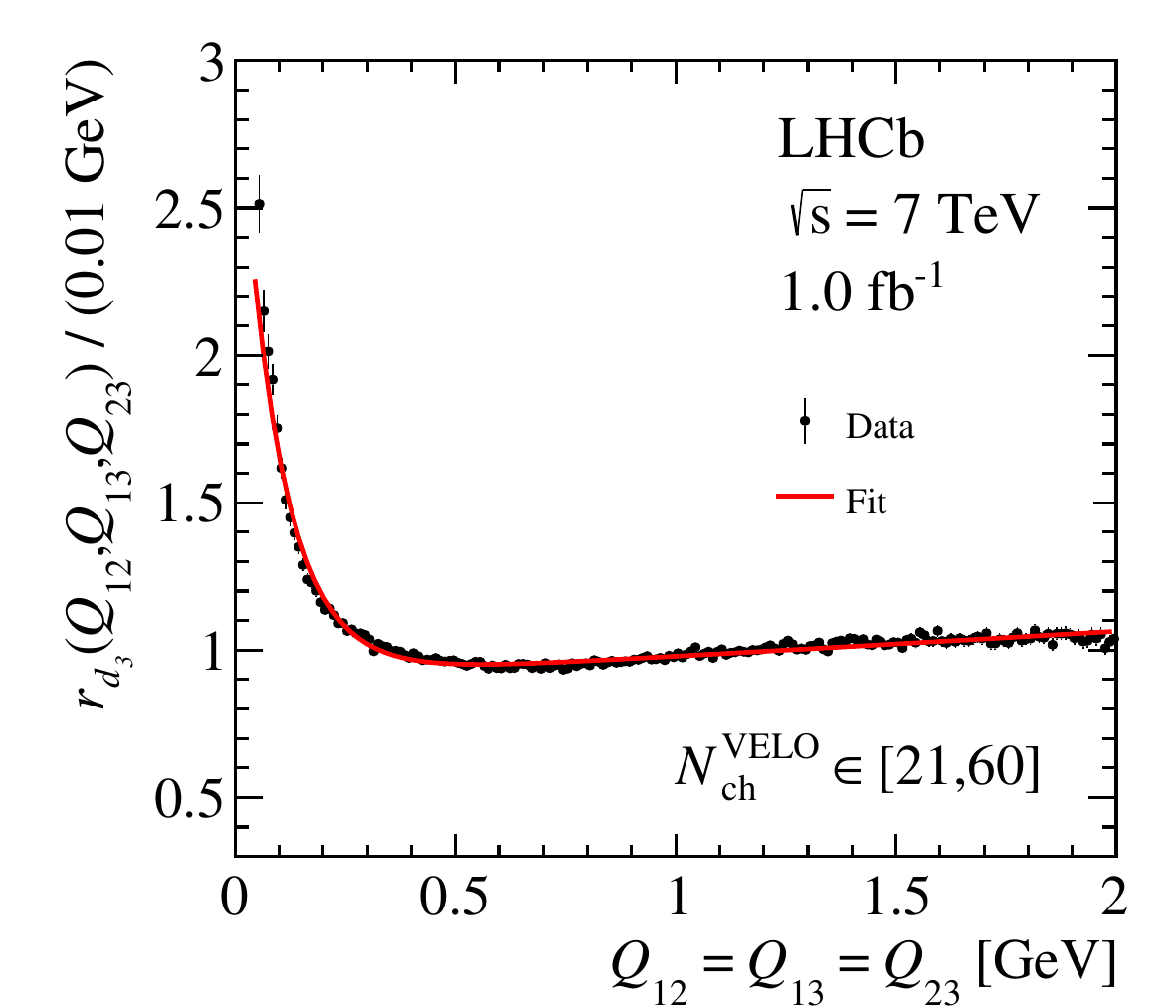}
    \end{center}
     \caption{Results of the fit to the three-particle double ratio ($r_{\rm d_{3}}$) for same-sign pion triplets, are presented in three bins of the VELO track multiplicity: (top left) [5,10], (top right) [11,20], (bottom) [21,60]. The red line illustrates the fit performed using the parameterisation outlined in Eq.~\ref{eq:3BECfullparam}, selecting the triplets of same-sign  pions with $Q$ values within the same bin.}
     \label{fig:3BEC_fits}
\end{figure}

\section{Systematic uncertainties}
\label{sec:systematics}

Several sources of systematic uncertainties are considered. The values for each source are summarised in Table~\ref{tab:systematic}, where the correlation functions are refitted using different requirements, and the difference with respect to the baseline result is assigned as a systematic uncertainty, excluding those found to be negligible. The total systematic uncertainty is obtained as the sum in quadrature of the individual components.

\begin{table}
    \begin{center}
        \caption{\label{tab:systematic}Systematic uncertainties split by sources and bins of VELO track multiplicity. Sources that proved to be negligible, not listed in the table, are mentioned in Sec.~\ref{sec:systematics}.}
        \begin{tabular}{l|c|c|c|c}
            \multicolumn{5}{c}{$N^{\text{VELO}}_{\text{ch}} \in $ [5,10]} \\
            \hline
            Source & $\sigma_{\lambda_3}$ [\%] & $\sigma_{f_c}$ [\%] & $\sigma_{p_c}$ [\%] & $\sigma_{\kappa_3}$ [\%] \\
            \hline
            Event generator & 8.7 & 10.4 & 9.7 & 10.4 \\
            PV multiplicity & 5.9 & 8.6 & 8.3 & 4.7 \\
            PV reconstruction & $< 0.1$ & 0.1 & 0.1 & $< 0.1$ \\
            Fit binning & 0.7 & 0.1 & 0.4 & 1.2 \\
            Fit low-$Q$ range & 0.3 & 0.6 & 1.0 & 0.3 \\
            Fit high-$Q$ range & 0.4 & 0.5 & 0.5 & 0.3 \\
            Fake tracks & 2.1 & 2.4 & 2.4 & 1.8 \\
            $\cal{P}(\pion)$ & 2.1 & 2.7 & 1.8 & 3.3 \\
            \hline
            Total & 11.0 & 14.0 & 13.2 & 12.1 \\
            \hline
            \multicolumn{5}{c}{}\\
            \multicolumn{5}{c}{$N^{\text{VELO}}_{\text{ch}} \in $ [11,20]} \\
            \hline
            Source & $\sigma_{\lambda_3}$ [\%] & $\sigma_{f_c}$ [\%] & $\sigma_{p_c}$ [\%] & $\sigma_{\kappa_3}$ [\%] \\
            \hline
            Event generator & 3.6 & 5.1 & 5.0 & 3.2 \\
            PV multiplicity & 2.1 & 5.0 & 5.0 & 2.1 \\
            PV reconstruction & 1.1 & 1.6 & 1.6 & 1.1 \\
            Fit binning & 1.1 & 1.1 & 1.1 & 1.1 \\
            Fit low-$Q$ range & 1.4 & 1.5 & 1.8 & 2.5 \\
            Fit high-$Q$ range & 1.8 & 1.9 & 1.8 & 2.1 \\
            Fake tracks & 1.8 & 4.3 & 4.3 & 1.8 \\
            $\cal{P}(\pion)$ & 2.1 & 1.7 & 1.4 & 3.9 \\
            \hline
            Total & 5.7 & 9.0 & 9.0 & 6.8 \\
            \hline
            \multicolumn{5}{c}{}\\
            \multicolumn{5}{c}{$N^{\text{VELO}}_{\text{ch}} \in $ [21,60]} \\
            \hline
            Source & $\sigma_{\lambda_3}$ [\%] & $\sigma_{f_c}$ [\%] & $\sigma_{p_c}$ [\%] & $\sigma_{\kappa_3}$ [\%] \\
            \hline
            Event generator & 2.4 & 2.8 & 2.8 & 0.4 \\
            PV multiplicity & 3.3 & 5.3 & 5.4 & 3.7 \\
            PV reconstruction & 2.4 & 2.5 & 2.4 & 2.8 \\
            Fit binning & 1.0 & 0.4 & 0.7 & 2.0 \\
            Fit low-$Q$ range & 1.8 & 2.4 & 3.0 & 3.7 \\
            Fit high-$Q$ range & 3.0 & 3.4 & 3.4 & 3.3 \\
            Fake tracks & 2.0 & 2.1 & 2.2 & 2.4 \\
            $\cal{P}(\pion)$ & 4.1 & 3.9 & 3.7 & 5.7 \\
            \hline
            Total & 7.5 & 8.9 & 9.1 & 9.4 \\
            \hline
        \end{tabular}
    \end{center}
\end{table}

The primary source of systematic uncertainty arises from differences in the event generators used to determine the correlation function in the simulation. To assess this effect, a sample of minimum-bias events generated with the \pythia 6.4 generator and the Perugia0 tune is used to construct the double ratio. The contribution to the systematic uncertainty is determined by comparing the central values of the results derived from the \pythia 8 and \pythia 6.4 datasets.

In the case of multiple primary vertices in a single event, the correlation function could be affected by the way the reference sample is formed. Residual correlations between primary vertices of a single event could be preserved and included in the correlation function. This contribution to the systematic uncertainty is calculated as a difference between the central values of the most differing fit results obtained using  two extreme cases: a single PV, two PVs, and three or four PVs in the event. For each subsample, a fit is performed, and the maximum difference observed for each measured parameter is taken as the systematic uncertainty. 

The systematic uncertainty due to the primary-vertex  reconstruction efficiency is also considered. To account for the effects of pile-up and inefficiencies in PV reconstruction, a systematic uncertainty is estimated by comparing the baseline fit results with those obtained from a fit to the data where the PV reconstruction is repeated after randomly removing a subset of tracks from the event.

Most fake tracks are already removed through the requirements on the track $\chi^2$ and the probability of being a fake track, as well as by applying a requirement on shared VELO hits, but there may be a discrepancy in fake-track ratios between the data and simulation. To determine the systematic uncertainty related to the fake tracks, the double ratio is refitted with a looser requirement on the probability of being a fake track $< 0.50$, the loosest possible value after preselection. The fraction of fake tracks after the final selection for three different bins of VELO track multiplicity and for different requirements on the probability of being a fake track $< 0.25$ and $< 0.50$ is less than 1\%. The fraction of same-sign pion pairs containing a clone track after the selection is found to be below 1\%. The systematic uncertainty arising from cloned tracks is estimated by fitting the double ratio after applying a strict requirement on the Kullback--Leibler distance~\cite{Needham:1082460}, ensuring that the clone contribution is completely removed in the simulation. The effect is determined to be negligible for all activity classes.

A requirement on $\cal{P}(\pion)$ reduces the contamination of misidentified pions. The systematic impact of this selection on the measured parameters is assessed using a looser requirement that increases the fraction of misidentified pions by~50\%. The systematic uncertainty due to the calibration of PID in the simulation is estimated by comparing different variants of the calibration procedure, using acceptances evaluated across various binning schemes for particle momentum, pseudorapidity, and track multiplicity. The largest difference observed after refitting the double ratios is taken as the systematic uncertainty.

The systematic uncertainty arising from the fit range in the low-$Q$ (high-$Q$) region is determined by adjusting the lower (upper) boundary of the $Q$ value by $\pm 0.01\gev$ \mbox{($\pm 0.20\gev$)}. Fits to the double ratios are performed for the three activity classes using two different lower (upper) $Q$ limits, and the largest difference observed is taken as a systematic uncertainty.

Other effects related to the Coulomb correction, fit binning, resolution of the $Q$ variable, different magnet polarities, beam-gas interactions and residual acceptance effects related to possible differences between data and simulation in the low-$Q$ region below 0.2\gev, are also investigated, and are found to be negligible.

\section{Results}
\label{results}

Fitted BEC parameters are used to calculate the values of the parameters in the core-halo model. The final results are presented in Table~\ref{tab:finalresults}, categorised into three bins based on VELO charged-particle multiplicity, along with their statistical and systematic uncertainties. As there is a correlation between the parameters $\ell_2$ and $\ell_3$, the uncertainty on $\lambda_{3}$ is determined taking into account this correlation, having however a negligible impact on the statistical uncertainty. Appendix~\ref{corr} includes fit parameters together with their correlation matrix.

\begin{table}
    \begin{center}
        \caption{\label{tab:finalresults} Measured values of core-halo parameters with statistical and systematic uncertainties, for different activity classes.}
        \begin{tabular}{c}
            \multicolumn{1}{c}{$N^{\text{VELO}}_{\text{ch}} \in $ [5,10]} \\
            \hline
            $\lambda_3$ = $3.37 \pm 0.24 \pm 0.37$ \\
            $f_c$ = $0.85 \pm 0.06 \pm 0.12$ \\
            $p_c$ = $0.08 \pm 0.01 \pm 0.01$ \\
            $\kappa_3$ = $0.99 \pm 0.09 \pm 0.12$ \\
            \hline
            \multicolumn{1}{c}{}\\
            \multicolumn{1}{c}{$N^{\text{VELO}}_{\text{ch}} \in $ [11,20]} \\
            \hline
            $\lambda_3$ = $2.80 \pm 0.17 \pm 0.16$ \\
            $f_c$ = $0.83 \pm 0.05 \pm 0.07$ \\
            $p_c$ = $0.29 \pm 0.02 \pm 0.03$ \\
            $\kappa_3$ = $0.91 \pm 0.09 \pm 0.06$ \\
            \hline
            \multicolumn{1}{c}{}\\
            \multicolumn{1}{c}{$N^{\text{VELO}}_{\text{ch}} \in $ [21,60]} \\
            \hline
            $\lambda_3$ = $2.46 \pm 0.20 \pm 0.18$ \\
            $f_c$ = $0.81 \pm 0.07 \pm 0.07$ \\
            $p_c$ = $0.35 \pm 0.03 \pm 0.03$ \\
            $\kappa_3$ = $0.87 \pm 0.12 \pm 0.08$ \\
            \hline
        \end{tabular}
    \end{center}
\end{table}

\begin{figure}
    \begin{center}
        \includegraphics[width=0.49\textwidth]{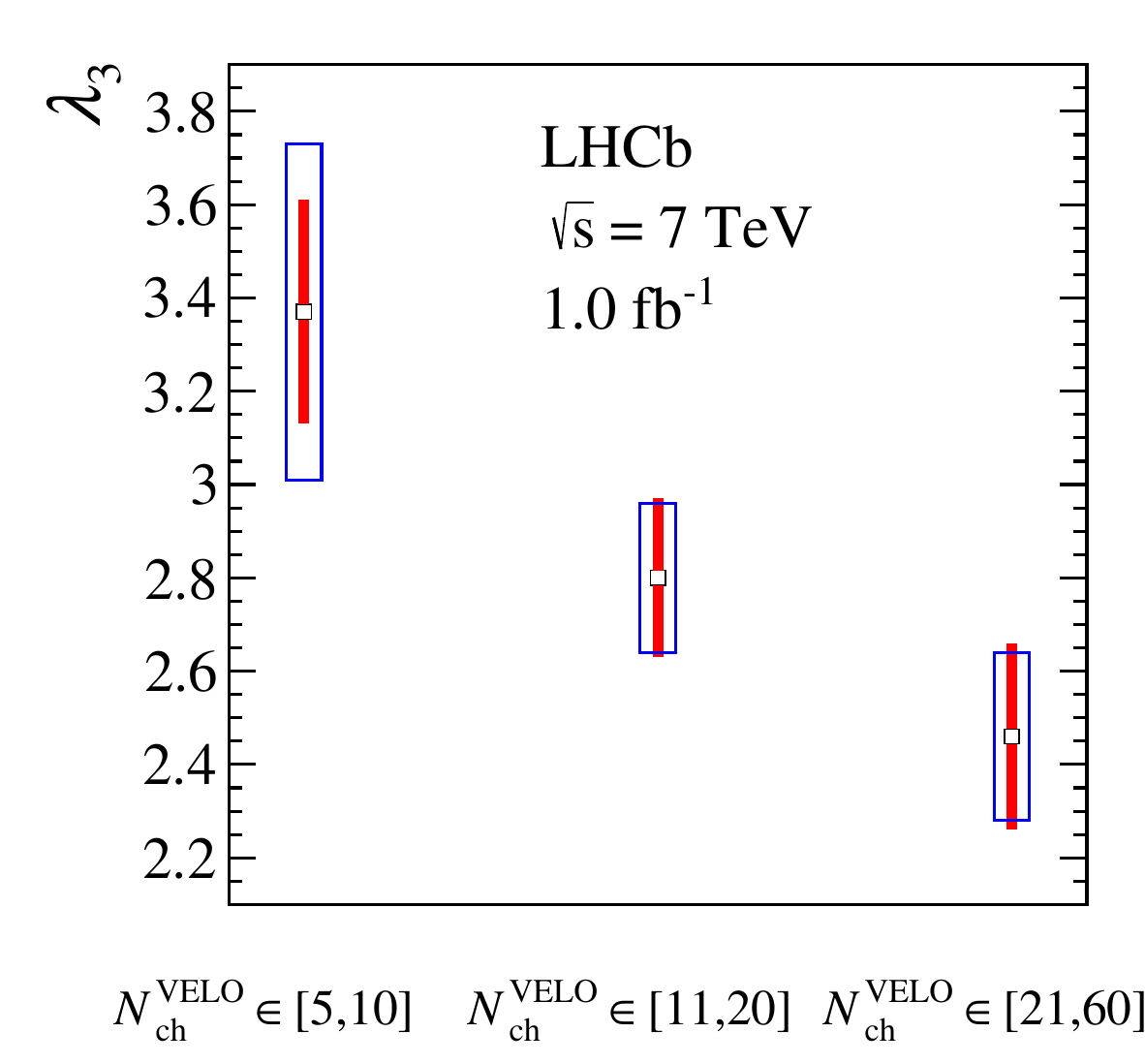} \hfill
        \includegraphics[width=0.49\textwidth]{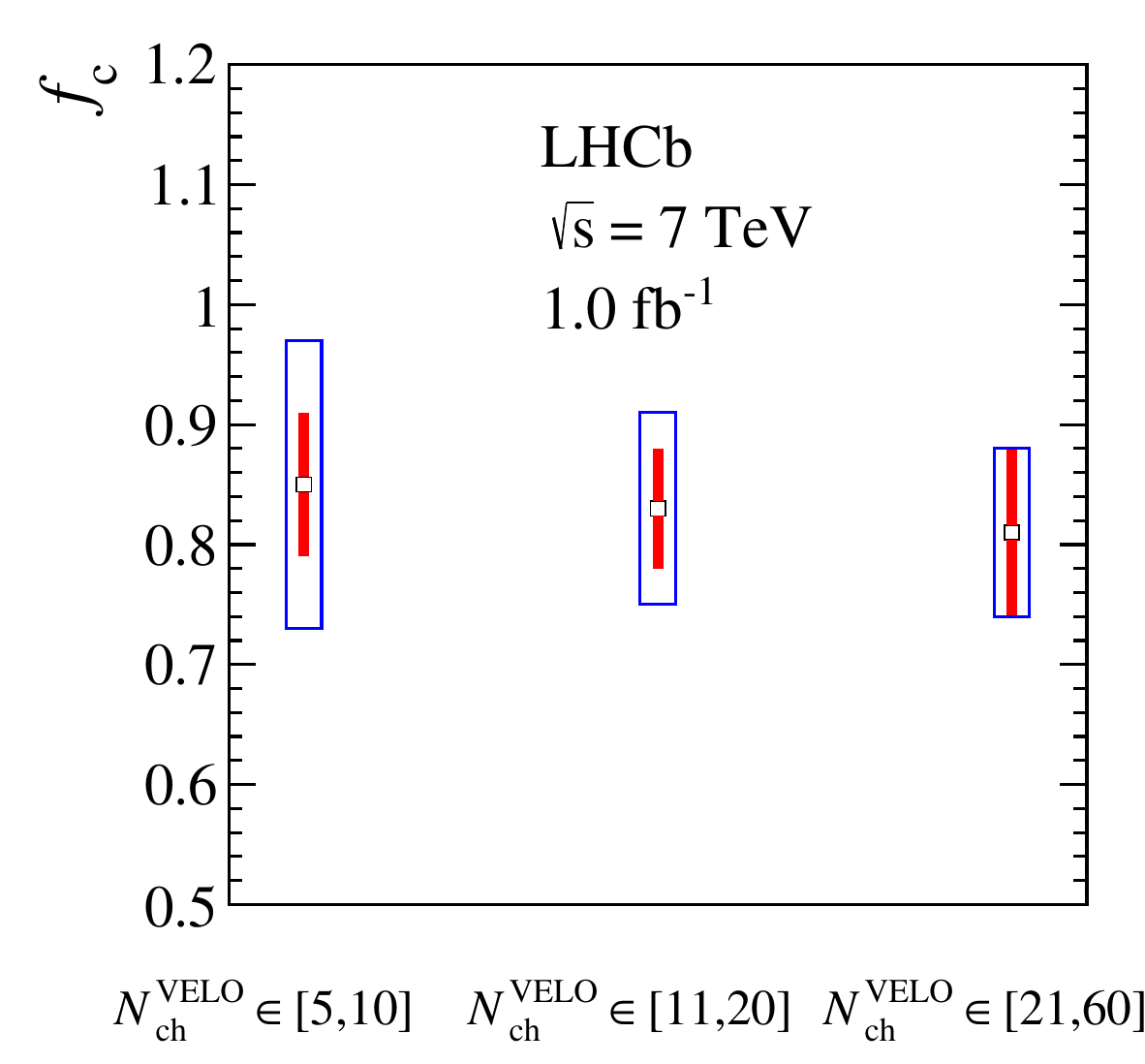} \\
        \includegraphics[width=0.49\textwidth]{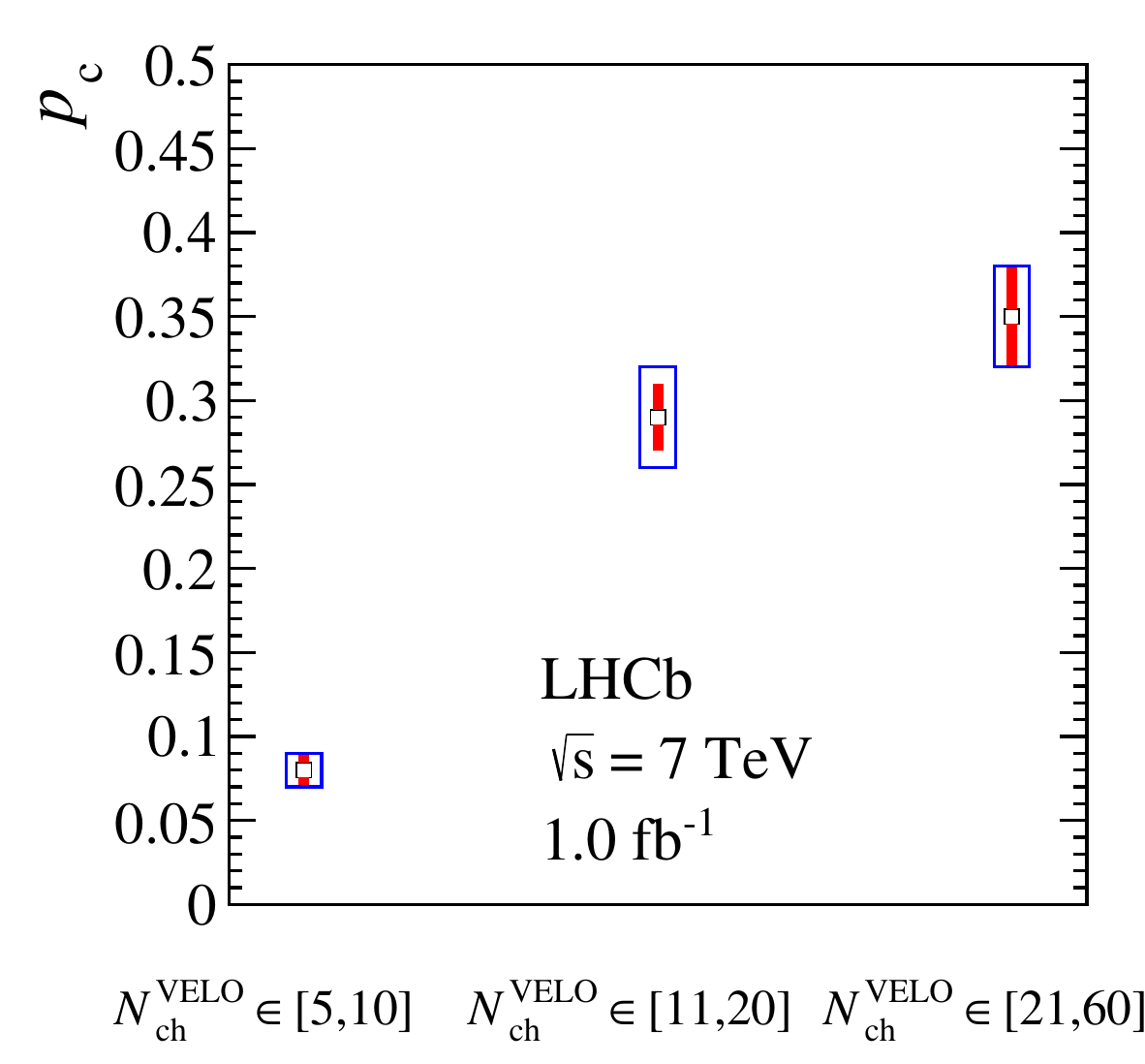} \hfill
        \includegraphics[width=0.49\textwidth]{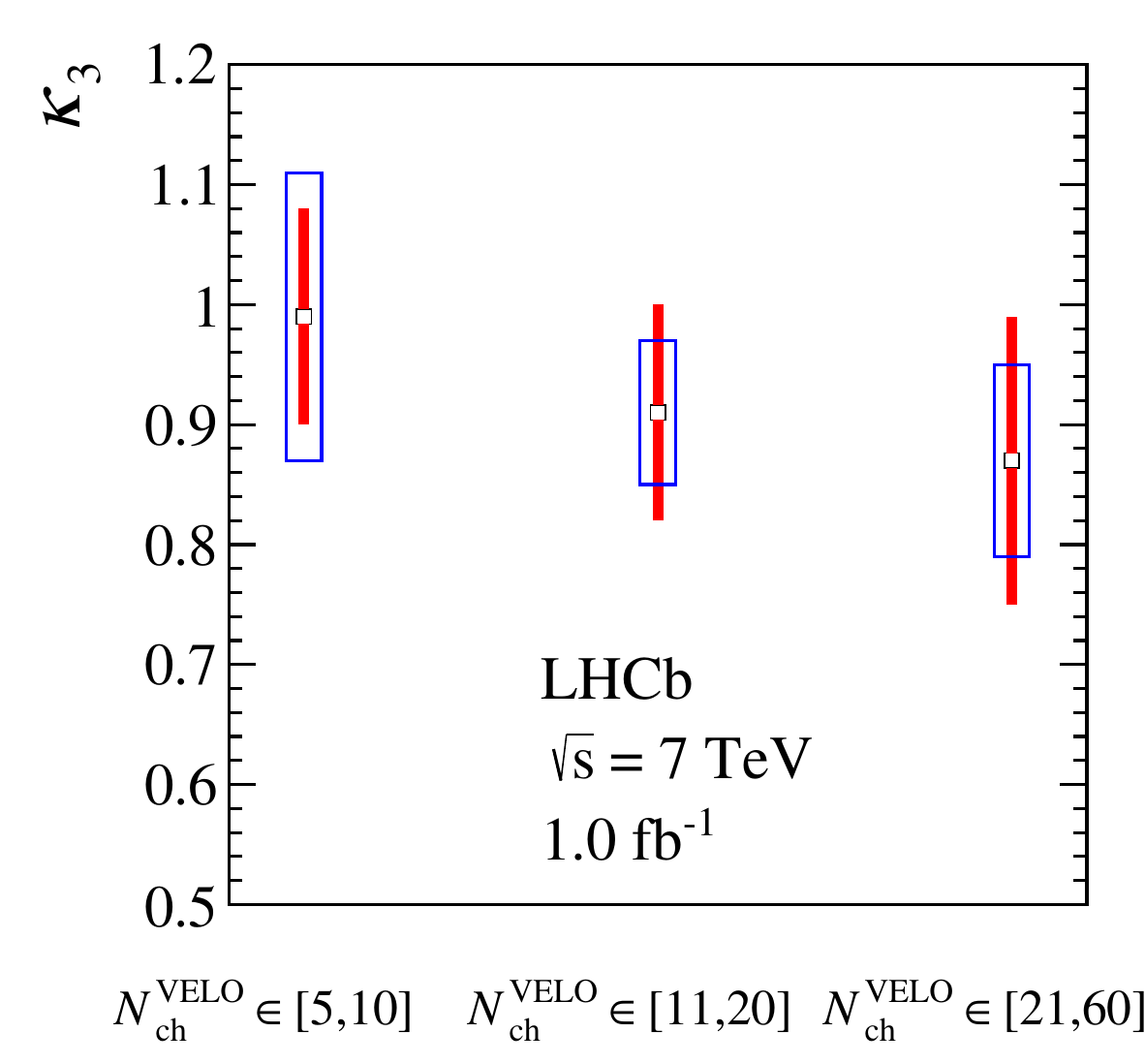}
    \end{center}
    \caption{Values of correlation strength $\lambda_3$ and the parameters of the core-halo model determined for three bins of VELO track multiplicity. Statistical and systematic uncertainties are marked with red and blue bars, respectively.}
    \label{fig:3BEC_results}
\end{figure}

The dependence of the core-halo parameters on the VELO track multiplicity is shown in Fig.~\ref{fig:3BEC_results}. The fraction of the core $f_c$ decreases only slightly for classes of higher activity, indicating that the fraction of particles that originated from the core of the emission volume does not depend strongly on the number of charged particles produced in the event. Conversely, the partial coherence $p_c$ increases significantly with higher multiplicities, showing the presence of partially coherent emission of pions. The central values of the parameter $\kappa_3$, which are systematically lower than unity for higher-multiplicity bins, also suggest a partially coherent emission. However, since they remain consistent with unity within statistical and systematic uncertainties, no definitive conclusion can be drawn. The parameter $\lambda_{3}$, related to the three-particle correlation strength is decreasing with charged-particle multiplicity, as expected. The systematic uncertainties are comparable to the statistical ones.

The analysis most closely aligned in scope with the present study was performed by the PHENIX experiment at RHIC~\cite{Phenix3BEC,art13}. It uses the core-halo model to interpret BEC for triplets of pions in Au-Au collisions at $\sqsnn = 200\gev$ in bins of transverse mass,\footnote{Transverse mass $m_{\text{T}} = \sqrt{m^2 + k_{\text{T}}^2}$, where $k_{\text{T}}$ is the average pair transverse momentum.} measuring $\kappa_3$ and correlation strength $\lambda_3$. Despite the differences in the experimental setups, the central values of $\kappa_3$ display a similar tendency to the results of the present study, which could indicate a partially coherent emission. However, these central values are consistent with unity within statistical and systematic uncertainties, preventing definitive conclusions. The reported values of three-particle correlation strengths are higher than those measured for \pp collisions in the present analysis. The PHENIX analysis does not present the dependence on particle multiplicity, therefore, a direct comparison of the results for \pp and Au-Au is not possible.

\section{Conclusions}
\label{conclusions}

The correlations among three indistinguishable pions are studied using data collected by the LHCb experiment in 2011 at a centre-of-mass energy of 7\tev in proton-proton collisions. The double-ratio technique is employed to correct for nonfemtoscopic background, together with the Gamov penetration factor utilised to account for electromagnetic final-state interactions that can be factorised using the generalised Riverside method. The fits applied to the three-pion double ratios of the correlation function provide parameters used to interpret the results within the frame of the core-halo model. The values of the core-halo parameters are studied in three activity classes related to the charged-particle multiplicity. The measured dependencies of the core-halo parameters on the charged-particle multiplicity suggest a partially coherent emission of pions. In particular, the partial coherence parameter $p_c$ increases with multiplicity, indicating coherent emission of pions. At the same time, the $\kappa_3$ parameter, which is systematically below unity for higher-multiplicity bins, further suggests a partially coherent emission. However, due to the large uncertainties, it is also consistent with unity, preventing any firm conclusions from being drawn in this case. The behaviour of $f_c$ indicates that the fraction of particles that originated from the core of the emission volume does not depend strongly on the number of charged particles produced in the event.

\newpage

\appendix
\section{Appendix. Fit parameters and correlation matrices}
\label{corr}

Parameters of the fit to the three-particle double ratio are summarised in Table~\ref{tab:fitresults}, while the correlation matrices of fitted parameters can be found in Tables~\ref{tab:correlationmatrix1}--\ref{tab:correlationmatrix3}.

\begin{table}[h]
    \begin{center}
        \caption{\label{tab:fitresults} The parameters of the fit to the three-particle double ratio.}
        \begin{tabular}{c|c|c|c}
             & $N^{\text{VELO}}_{\text{ch}} \in $ [5,10] & $N^{\text{VELO}}_{\text{ch}} \in $ [11,20] & $N^{\text{VELO}}_{\text{ch}} \in $ [21,60]\\
            \hline
            $\chi^{2}/ndf$ & $170/191$ & $285/191$ & $262/191$ \\
            $N$ & $0.74 \pm 0.01$ & $0.85 \pm 0.01$ & $0.90 \pm 0.01$\\
            $\ell_2$ & $0.47 \pm 0.05$ & $0.52 \pm 0.03$ & $0.46 \pm 0.05$ \\
            $\ell_3$ & $1.93 \pm 0.19$ & $1.25 \pm 0.15$ & $1.08 \pm 0.18$ \\
            $\delta$ & $0.07 \pm 0.01$ & $0.04 \pm 0.01$ & $0.03 \pm 0.01$ \\
            \hline
            
        \end{tabular}
    \end{center}
\end{table}

\begin{table}[h]
    \begin{center}
        \caption{\label{tab:correlationmatrix1}Correlation matrix between fit parameters for $N^{\text{VELO}}_{\text{ch}} \in $ [5,10].}
        \begin{tabular}{c|c|c|c|c}
            & $N$ & $\delta$ & $\ell_3$ & $\ell_2$ \\
            \hline
            $N$ & $1$ & $-0.97$ & \phantom{$-$}$0.66$ & $-0.85$ \\
            $\delta$ & $-0.97$ & $1$ & $-0.58$ & \phantom{$-$}$0.77$ \\
            $\ell_3$ & \phantom{$-$}$0.66$ & $-0.58$ & $1$ & $-0.94$ \\
            $\ell_2$ & $-0.85$ & \phantom{$-$}$0.77$ & $-0.94$ & $1$ \\
            \hline
        \end{tabular}
  \end{center}
\end{table}

\begin{table}[h]
    \begin{center}
        \caption{\label{tab:correlationmatrix2}Correlation matrix between fit parameters for $N^{\text{VELO}}_{\text{ch}} \in $ [11,20].}
        \begin{tabular}{c|c|c|c|c}
            & $N$ & $\delta$ & $\ell_3$ & $\ell_2$ \\
            \hline
            $N$ & $1$ & $-0.95$ & \phantom{$-$}$0.57$ & $-0.73$ \\
            $\delta$ & $-0.95$ & $1$ & $-0.49$ & \phantom{$-$}$0.64$ \\
            $\ell_3$ & \phantom{$-$}$0.57$ & $-0.49$ & $1$ & $-0.95$ \\
            $\ell_2$ & $-0.73$ & \phantom{$-$}$0.64$ & $-0.95$ & $1$ \\
            \hline
        \end{tabular}
  \end{center}
\end{table}

\begin{table}[h]
    \begin{center}
        \caption{\label{tab:correlationmatrix3}Correlation matrix between fit parameters for $N^{\text{VELO}}_{\text{ch}} \in $ [21,60].}
        \begin{tabular}{c|c|c|c|c}
            & $N$ & $\delta$ & $\ell_3$ & $\ell_2$ \\
            \hline
            $N$ & $1$ & $-0.94$ & \phantom{$-$}$0.50$ & $-0.64$ \\
            $\delta$ & $-0.94$ & $1$ & $-0.43$ & \phantom{$-$}$0.56$ \\
            $\ell_3$ & \phantom{$-$}$0.50$ & $-0.43$ & $1$ & $-0.95$ \\
            $\ell_2$ & $-0.64$ & \phantom{$-$}$0.56$ & $-0.95$ & $1$ \\
            \hline
        \end{tabular}
  \end{center}
\end{table}


\newpage
\section*{Acknowledgements}
%
%
\noindent We express our gratitude to our colleagues in the CERN
accelerator departments for the excellent performance of the LHC. We
thank the technical and administrative staff at the LHCb
institutes.
We acknowledge support from CERN and from the national agencies:
ARC (Australia);
CAPES, CNPq, FAPERJ and FINEP (Brazil); 
MOST and NSFC (China); 
CNRS/IN2P3 (France); 
BMBF, DFG and MPG (Germany); 
INFN (Italy); 
NWO (Netherlands); 
MNiSW and NCN (Poland); 
MCID/IFA (Romania); 
MICIU and AEI (Spain);
SNSF and SER (Switzerland); 
NASU (Ukraine); 
STFC (United Kingdom); 
DOE NP and NSF (USA).
We acknowledge the computing resources that are provided by ARDC (Australia), 
CBPF (Brazil),
CERN, 
IHEP and LZU (China),
IN2P3 (France), 
KIT and DESY (Germany), 
INFN (Italy), 
SURF (Netherlands),
Polish WLCG (Poland),
IFIN-HH (Romania), 
PIC (Spain), CSCS (Switzerland), 
and GridPP (United Kingdom).
We are indebted to the communities behind the multiple open-source
software packages on which we depend.
Individual groups or members have received support from
Key Research Program of Frontier Sciences of CAS, CAS PIFI, CAS CCEPP, 
Fundamental Research Funds for the Central Universities,  and Sci.\ \& Tech.\ Program of Guangzhou (China);
Minciencias (Colombia);
EPLANET, Marie Sk\l{}odowska-Curie Actions, ERC and NextGenerationEU (European Union);
A*MIDEX, ANR, IPhU and Labex P2IO, and R\'{e}gion Auvergne-Rh\^{o}ne-Alpes (France);
Alexander-von-Humboldt Foundation (Germany);
ICSC (Italy); 
Severo Ochoa and Mar\'ia de Maeztu Units of Excellence, GVA, XuntaGal, GENCAT, InTalent-Inditex and Prog.~Atracci\'on Talento CM (Spain);
SRC (Sweden);
the Leverhulme Trust, the Royal Society and UKRI (United Kingdom).




\newpage
\addcontentsline{toc}{section}{References}
\bibliographystyle{LHCb}
\bibliography{main,standard,LHCb-PAPER,LHCb-CONF,LHCb-DP,LHCb-TDR}

\ifx\mcitethebibliography\mciteundefinedmacro
\PackageError{LHCb.bst}{mciteplus.sty has not been loaded}
{This bibstyle requires the use of the mciteplus package.}\fi
\providecommand{\href}[2]{#2}
\begin{mcitethebibliography}{10}
\mciteSetBstSublistMode{n}
\mciteSetBstMaxWidthForm{subitem}{\alph{mcitesubitemcount})}
\mciteSetBstSublistLabelBeginEnd{\mcitemaxwidthsubitemform\space}
{\relax}{\relax}

\bibitem{HanburyBrown:1954}
R.~Hanbury~Brown and R.~Q. Twiss, \ifthenelse{\boolean{articletitles}}{\emph{{A New type of interferometer for use in radio astronomy}}, }{}\href{https://doi.org/10.1080/14786440708520475}{Phil.\ Mag.\ Ser.\ 7 \textbf{45} (1954) 663}\relax
\mciteBstWouldAddEndPuncttrue
\mciteSetBstMidEndSepPunct{\mcitedefaultmidpunct}
{\mcitedefaultendpunct}{\mcitedefaultseppunct}\relax
\EndOfBibitem
\bibitem{Brown:1956}
R.~Hanbury~Brown and R.~Q. Twiss, \ifthenelse{\boolean{articletitles}}{\emph{{Correlation between photons in two coherent beams of light}}, }{}\href{https://doi.org/10.1038/177027a0}{Nature \textbf{177} (1956) 27}\relax
\mciteBstWouldAddEndPuncttrue
\mciteSetBstMidEndSepPunct{\mcitedefaultmidpunct}
{\mcitedefaultendpunct}{\mcitedefaultseppunct}\relax
\EndOfBibitem
\bibitem{HanburyBrown:1956}
R.~Hanbury~Brown and R.~Q. Twiss, \ifthenelse{\boolean{articletitles}}{\emph{{A test of a new type of stellar interferometer on Sirius}}, }{}\href{https://doi.org/10.1038/1781046a0}{Nature \textbf{178} (1956) 1046}\relax
\mciteBstWouldAddEndPuncttrue
\mciteSetBstMidEndSepPunct{\mcitedefaultmidpunct}
{\mcitedefaultendpunct}{\mcitedefaultseppunct}\relax
\EndOfBibitem
\bibitem{Lisa_2005}
M.~A. Lisa, S.~Pratt, R.~Soltz, and U.~Wiedemann, \ifthenelse{\boolean{articletitles}}{\emph{Femtoscopy in relativistic heavy ion collisions: Two decades of progress}, }{}\href{https://doi.org/10.1146/annurev.nucl.55.090704.151533}{Annu.\ Rev.\ Nucl.\ Sci.\  \textbf{55} (2005) 357}, \href{http://arxiv.org/abs/nucl-ex/0505014}{{\normalfont\ttfamily arXiv:nucl-ex/0505014}}\relax
\mciteBstWouldAddEndPuncttrue
\mciteSetBstMidEndSepPunct{\mcitedefaultmidpunct}
{\mcitedefaultendpunct}{\mcitedefaultseppunct}\relax
\EndOfBibitem
\bibitem{Schenke:2014}
B.~Schenke and R.~Venugopalan, \ifthenelse{\boolean{articletitles}}{\emph{{Eccentric protons? Sensitivity of flow to system size and shape in p+p, p+Pb and Pb+Pb collisions}}, }{}\href{https://doi.org/10.1103/PhysRevLett.113.102301}{Phys.\ Rev.\ Lett.\  \textbf{113} (2014) 102301}, \href{http://arxiv.org/abs/1405.3605}{{\normalfont\ttfamily arXiv:1405.3605}}\relax
\mciteBstWouldAddEndPuncttrue
\mciteSetBstMidEndSepPunct{\mcitedefaultmidpunct}
{\mcitedefaultendpunct}{\mcitedefaultseppunct}\relax
\EndOfBibitem
\bibitem{s10052-015-3509-3}
P.~Romatschke, \ifthenelse{\boolean{articletitles}}{\emph{{Light-heavy-ion collisions: a window into pre-equilibrium QCD dynamics?}}, }{}\href{https://doi.org/10.1140/epjc/s10052-015-3509-3}{Eur.\ Phys.\ J.\  \textbf{C75} (2015) 305}, \href{http://arxiv.org/abs/1502.04745}{{\normalfont\ttfamily arXiv:1502.04745}}\relax
\mciteBstWouldAddEndPuncttrue
\mciteSetBstMidEndSepPunct{\mcitedefaultmidpunct}
{\mcitedefaultendpunct}{\mcitedefaultseppunct}\relax
\EndOfBibitem
\bibitem{Bialas:2014gca}
A.~Bialas, W.~Florkowski, and K.~Zalewski, \ifthenelse{\boolean{articletitles}}{\emph{{Bose--Einstein correlations and thermal cluster formation in high-energy collisions}}, }{}\href{https://doi.org/10.5506/APhysPolB.45.1883}{Acta Phys.\ Polon.\  \textbf{B45} (2014) 1883}, \href{http://arxiv.org/abs/1406.2499}{{\normalfont\ttfamily arXiv:1406.2499}}\relax
\mciteBstWouldAddEndPuncttrue
\mciteSetBstMidEndSepPunct{\mcitedefaultmidpunct}
{\mcitedefaultendpunct}{\mcitedefaultseppunct}\relax
\EndOfBibitem
\bibitem{GOTTSCHALK1984325}
T.~D. Gottschalk, \ifthenelse{\boolean{articletitles}}{\emph{{A simple phenomenological model for hadron production from low-mass clusters}}, }{}\href{https://doi.org/https://doi.org/10.1016/0550-3213(84)90252-9}{Nucl.\ Phys.\  \textbf{B239} (1984) 325}\relax
\mciteBstWouldAddEndPuncttrue
\mciteSetBstMidEndSepPunct{\mcitedefaultmidpunct}
{\mcitedefaultendpunct}{\mcitedefaultseppunct}\relax
\EndOfBibitem
\bibitem{Csorgo:1994in}
T.~Cs\"org\H{o}, B.~L\"orstad, and J.~Zimányi, \ifthenelse{\boolean{articletitles}}{\emph{{Bose--Einstein correlations for systems with large halo}}, }{}\href{https://doi.org/10.1007/BF02907008}{Z.\ Phys.\  \textbf{C71} (1996) 491}, \href{http://arxiv.org/abs/hep-ph/9411307}{{\normalfont\ttfamily arXiv:hep-ph/9411307}}\relax
\mciteBstWouldAddEndPuncttrue
\mciteSetBstMidEndSepPunct{\mcitedefaultmidpunct}
{\mcitedefaultendpunct}{\mcitedefaultseppunct}\relax
\EndOfBibitem
\bibitem{PhysRevC.89.024911}
ALICE collaboration, B.~Abelev {\em et~al.}, \ifthenelse{\boolean{articletitles}}{\emph{{Two- and three-pion quantum statistics correlations in Pb-Pb collisions at $\sqsnn$=2.76\tev at the CERN Large Hadron Collider}}, }{}\href{https://doi.org/10.1103/PhysRevC.89.024911}{Phys.\ Rev.\  \textbf{C89} (2014) 024911}, \href{http://arxiv.org/abs/1310.7808}{{\normalfont\ttfamily arXiv:1310.7808}}\relax
\mciteBstWouldAddEndPuncttrue
\mciteSetBstMidEndSepPunct{\mcitedefaultmidpunct}
{\mcitedefaultendpunct}{\mcitedefaultseppunct}\relax
\EndOfBibitem
\bibitem{Phenix3BEC}
T.~Novák, \ifthenelse{\boolean{articletitles}}{\emph{{PHENIX results of three-particle Bose--Einstein correlations in \mbox{$\sqsnn = 200$\gev} Au+Au collisions}}, }{}\href{https://doi.org/10.3390/universe4030057}{Universe \textbf{4} (2018) 57}, \href{http://arxiv.org/abs/1801.03544}{{\normalfont\ttfamily arXiv:1801.03544}}\relax
\mciteBstWouldAddEndPuncttrue
\mciteSetBstMidEndSepPunct{\mcitedefaultmidpunct}
{\mcitedefaultendpunct}{\mcitedefaultseppunct}\relax
\EndOfBibitem
\bibitem{art13}
M.~Csan\'ad, \ifthenelse{\boolean{articletitles}}{\emph{{Two- and three-pion L\'evy femtoscopy with PHENIX}}, }{}\href{https://doi.org/10.1088/1742-6596/1070/1/012026}{J.\ Phys.\ : Conf.\ Ser.\  \textbf{1070} (2018) 012026}, \href{http://arxiv.org/abs/1806.05745}{{\normalfont\ttfamily arXiv:1806.05745}}\relax
\mciteBstWouldAddEndPuncttrue
\mciteSetBstMidEndSepPunct{\mcitedefaultmidpunct}
{\mcitedefaultendpunct}{\mcitedefaultseppunct}\relax
\EndOfBibitem
\bibitem{LHCb-PAPER-2017-025}
LHCb collaboration, R.~Aaij {\em et~al.}, \ifthenelse{\boolean{articletitles}}{\emph{{Bose--Einstein correlations of same-sign charged pions in the forward region in \proton\proton collisions at \mbox{$\sqs=$7~\tev}}}, }{}\href{https://doi.org/10.1007/JHEP12(2017)025}{JHEP \textbf{12} (2017) 025}, \href{http://arxiv.org/abs/1709.01769}{{\normalfont\ttfamily arXiv:1709.01769}}\relax
\mciteBstWouldAddEndPuncttrue
\mciteSetBstMidEndSepPunct{\mcitedefaultmidpunct}
{\mcitedefaultendpunct}{\mcitedefaultseppunct}\relax
\EndOfBibitem
\bibitem{Csorgo2000}
{Cs{\"o}rg{\H{o}}, T.\ }, \ifthenelse{\boolean{articletitles}}{\emph{{Particle interferometry from $40$\mev to $40$\tev}}, }{}\href{https://doi.org/10.1007/978-94-011-4126-0_8}{Nato Science Series C: Math.\ and Phys.\ Sci.\  \textbf{554} (2000) 203}, \href{http://arxiv.org/abs/hep-ph/0001233}{{\normalfont\ttfamily arXiv:hep-ph/0001233}}\relax
\mciteBstWouldAddEndPuncttrue
\mciteSetBstMidEndSepPunct{\mcitedefaultmidpunct}
{\mcitedefaultendpunct}{\mcitedefaultseppunct}\relax
\EndOfBibitem
\bibitem{Baym:1997}
G.~Baym, \ifthenelse{\boolean{articletitles}}{\emph{{The physics of Hanbury Brown--Twiss intensity interferometry: from stars to nuclear collisions}}, }{}Acta Phys.\ Polon.\  \textbf{B29} (1998) 1839, \href{http://arxiv.org/abs/nucl-th/9804026}{{\normalfont\ttfamily arXiv:nucl-th/9804026}}\relax
\mciteBstWouldAddEndPuncttrue
\mciteSetBstMidEndSepPunct{\mcitedefaultmidpunct}
{\mcitedefaultendpunct}{\mcitedefaultseppunct}\relax
\EndOfBibitem
\bibitem{Alexander:2003ug}
G.~Alexander, \ifthenelse{\boolean{articletitles}}{\emph{{Bose--Einstein and Fermi--Dirac interferometry in particle physics}}, }{}\href{https://doi.org/10.1088/0034-4885/66/4/202}{Rept.\ Prog.\ Phys.\  \textbf{66} (2003) 481}, \href{http://arxiv.org/abs/hep-ph/0302130}{{\normalfont\ttfamily arXiv:hep-ph/0302130}}\relax
\mciteBstWouldAddEndPuncttrue
\mciteSetBstMidEndSepPunct{\mcitedefaultmidpunct}
{\mcitedefaultendpunct}{\mcitedefaultseppunct}\relax
\EndOfBibitem
\bibitem{BECforLevyStable}
T.~Cs\"org\H{o}, S.~Hegyi, and W.~A. Zajc, \ifthenelse{\boolean{articletitles}}{\emph{{Bose--Einstein correlations for L\'evy stable source distributions}}, }{}\href{https://doi.org/10.1140/epjc/s2004-01870-9}{Eur.\ Phys.\ J.\  \textbf{C36} (2004) 67}, \href{http://arxiv.org/abs/nucl-th/0310042}{{\normalfont\ttfamily arXiv:nucl-th/0310042}}\relax
\mciteBstWouldAddEndPuncttrue
\mciteSetBstMidEndSepPunct{\mcitedefaultmidpunct}
{\mcitedefaultendpunct}{\mcitedefaultseppunct}\relax
\EndOfBibitem
\bibitem{Abbiendi:2000jb}
OPAL collaboration, G.~Abbiendi {\em et~al.}, \ifthenelse{\boolean{articletitles}}{\emph{{Bose--Einstein correlations in \Kpm\Kpm pairs from \Z decays into two hadronic jets}}, }{}\href{https://doi.org/10.1007/s100520100715}{Eur.\ Phys.\ J.\  \textbf{C21} (2001) 23}, \href{http://arxiv.org/abs/hep-ex/0001045}{{\normalfont\ttfamily arXiv:hep-ex/0001045}}\relax
\mciteBstWouldAddEndPuncttrue
\mciteSetBstMidEndSepPunct{\mcitedefaultmidpunct}
{\mcitedefaultendpunct}{\mcitedefaultseppunct}\relax
\EndOfBibitem
\bibitem{wil2}
S.~Pratt, T.~Cs{\"o}rg{\H{o}}, and J.~Zimányi, \ifthenelse{\boolean{articletitles}}{\emph{{Detailed predictions for two pion correlations in ultrarelativistic heavy ion collisions}}, }{}\href{https://doi.org/10.1103/PhysRevC.42.2646}{Phys.\ Rev.\  \textbf{C42} (1990) 2646}\relax
\mciteBstWouldAddEndPuncttrue
\mciteSetBstMidEndSepPunct{\mcitedefaultmidpunct}
{\mcitedefaultendpunct}{\mcitedefaultseppunct}\relax
\EndOfBibitem
\bibitem{wil3}
S.~Chapman and U.~W. Heinz, \ifthenelse{\boolean{articletitles}}{\emph{{HBT correlators - current formalism vs. Wigner function formulation}}, }{}\href{https://doi.org/10.1016/0370-2693(94)01277-6}{Phys.\ Lett.\  \textbf{B340} (1994) 250}\relax
\mciteBstWouldAddEndPuncttrue
\mciteSetBstMidEndSepPunct{\mcitedefaultmidpunct}
{\mcitedefaultendpunct}{\mcitedefaultseppunct}\relax
\EndOfBibitem
\bibitem{PhysRevC.20.2267}
M.~Gyulassy, S.~K. Kauffmann, and L.~W. Wilson, \ifthenelse{\boolean{articletitles}}{\emph{{Pion interferometry of nuclear collisions}}, }{}\href{https://doi.org/10.1103/PhysRevC.20.2267}{Phys.\ Rev.\  \textbf{C20} (1979) 2267}\relax
\mciteBstWouldAddEndPuncttrue
\mciteSetBstMidEndSepPunct{\mcitedefaultmidpunct}
{\mcitedefaultendpunct}{\mcitedefaultseppunct}\relax
\EndOfBibitem
\bibitem{riverside}
D.~Gangadharan, \ifthenelse{\boolean{articletitles}}{\emph{{Techniques for multiboson interferometry}}, }{}\href{https://doi.org/10.1103/PhysRevC.92.014902}{Phys.\ Rev.\  \textbf{C92} (2015) 014902}, \href{http://arxiv.org/abs/1502.02121}{{\normalfont\ttfamily arXiv:1502.02121}}\relax
\mciteBstWouldAddEndPuncttrue
\mciteSetBstMidEndSepPunct{\mcitedefaultmidpunct}
{\mcitedefaultendpunct}{\mcitedefaultseppunct}\relax
\EndOfBibitem
\bibitem{PhysRevD.33.72}
S.~Pratt, \ifthenelse{\boolean{articletitles}}{\emph{{Coherence and Coulomb effects on pion interferometry}}, }{}\href{https://doi.org/10.1103/PhysRevD.33.72}{Phys.\ Rev.\  \textbf{D33} (1986) 72}\relax
\mciteBstWouldAddEndPuncttrue
\mciteSetBstMidEndSepPunct{\mcitedefaultmidpunct}
{\mcitedefaultendpunct}{\mcitedefaultseppunct}\relax
\EndOfBibitem
\bibitem{Sirunyan:2017}
CMS collaboration, A.~M. Sirunyan {\em et~al.}, \ifthenelse{\boolean{articletitles}}{\emph{{Bose--Einstein correlations in $pp$, $p$Pb, and PbPb collisions at $\sqsnn= 0.9$-$7$\tev}}, }{}\href{https://doi.org/10.1103/PhysRevC.97.064912}{Phys.\ Rev.\  \textbf{C97} (2018) 064912}, \href{http://arxiv.org/abs/1712.07198}{{\normalfont\ttfamily arXiv:1712.07198}}\relax
\mciteBstWouldAddEndPuncttrue
\mciteSetBstMidEndSepPunct{\mcitedefaultmidpunct}
{\mcitedefaultendpunct}{\mcitedefaultseppunct}\relax
\EndOfBibitem
\bibitem{riverside2}
Y.~M. Liu {\em et~al.}, \ifthenelse{\boolean{articletitles}}{\emph{{Three-pion correlations in relativistic heavy ion collisions}}, }{}\href{https://doi.org/10.1103/PhysRevC.34.1667}{Phys.\ Rev.\  \textbf{C34} (1986) 1667}\relax
\mciteBstWouldAddEndPuncttrue
\mciteSetBstMidEndSepPunct{\mcitedefaultmidpunct}
{\mcitedefaultendpunct}{\mcitedefaultseppunct}\relax
\EndOfBibitem
\bibitem{LHCb-DP-2008-001}
LHCb collaboration, A.~A. Alves~Jr.\ {\em et~al.}, \ifthenelse{\boolean{articletitles}}{\emph{{The \lhcb detector at the LHC}}, }{}\href{https://doi.org/10.1088/1748-0221/3/08/S08005}{JINST \textbf{3} (2008) S08005}\relax
\mciteBstWouldAddEndPuncttrue
\mciteSetBstMidEndSepPunct{\mcitedefaultmidpunct}
{\mcitedefaultendpunct}{\mcitedefaultseppunct}\relax
\EndOfBibitem
\bibitem{LHCb-DP-2014-002}
LHCb collaboration, R.~Aaij {\em et~al.}, \ifthenelse{\boolean{articletitles}}{\emph{{LHCb detector performance}}, }{}\href{https://doi.org/10.1142/S0217751X15300227}{Int.\ J.\ Mod.\ Phys.\  \textbf{A30} (2015) 1530022}, \href{http://arxiv.org/abs/1412.6352}{{\normalfont\ttfamily arXiv:1412.6352}}\relax
\mciteBstWouldAddEndPuncttrue
\mciteSetBstMidEndSepPunct{\mcitedefaultmidpunct}
{\mcitedefaultendpunct}{\mcitedefaultseppunct}\relax
\EndOfBibitem
\bibitem{LHCb-DP-2014-001}
R.~Aaij {\em et~al.}, \ifthenelse{\boolean{articletitles}}{\emph{{Performance of the LHCb Vertex Locator}}, }{}\href{https://doi.org/10.1088/1748-0221/9/09/P09007}{JINST \textbf{9} (2014) P09007}, \href{http://arxiv.org/abs/1405.7808}{{\normalfont\ttfamily arXiv:1405.7808}}\relax
\mciteBstWouldAddEndPuncttrue
\mciteSetBstMidEndSepPunct{\mcitedefaultmidpunct}
{\mcitedefaultendpunct}{\mcitedefaultseppunct}\relax
\EndOfBibitem
\bibitem{LHCb-DP-2013-003}
R.~Arink {\em et~al.}, \ifthenelse{\boolean{articletitles}}{\emph{{Performance of the LHCb Outer Tracker}}, }{}\href{https://doi.org/10.1088/1748-0221/9/01/P01002}{JINST \textbf{9} (2014) P01002}, \href{http://arxiv.org/abs/1311.3893}{{\normalfont\ttfamily arXiv:1311.3893}}\relax
\mciteBstWouldAddEndPuncttrue
\mciteSetBstMidEndSepPunct{\mcitedefaultmidpunct}
{\mcitedefaultendpunct}{\mcitedefaultseppunct}\relax
\EndOfBibitem
\bibitem{LHCb-DP-2012-003}
M.~Adinolfi {\em et~al.}, \ifthenelse{\boolean{articletitles}}{\emph{{Performance of the \lhcb RICH detector at the LHC}}, }{}\href{https://doi.org/10.1140/epjc/s10052-013-2431-9}{Eur.\ Phys.\ J.\  \textbf{C73} (2013) 2431}, \href{http://arxiv.org/abs/1211.6759}{{\normalfont\ttfamily arXiv:1211.6759}}\relax
\mciteBstWouldAddEndPuncttrue
\mciteSetBstMidEndSepPunct{\mcitedefaultmidpunct}
{\mcitedefaultendpunct}{\mcitedefaultseppunct}\relax
\EndOfBibitem
\bibitem{LHCb-DP-2012-002}
A.~A. Alves~Jr.\ {\em et~al.}, \ifthenelse{\boolean{articletitles}}{\emph{{Performance of the LHCb muon system}}, }{}\href{https://doi.org/10.1088/1748-0221/8/02/P02022}{JINST \textbf{8} (2013) P02022}, \href{http://arxiv.org/abs/1211.1346}{{\normalfont\ttfamily arXiv:1211.1346}}\relax
\mciteBstWouldAddEndPuncttrue
\mciteSetBstMidEndSepPunct{\mcitedefaultmidpunct}
{\mcitedefaultendpunct}{\mcitedefaultseppunct}\relax
\EndOfBibitem
\bibitem{LHCb-DP-2012-004}
R.~Aaij {\em et~al.}, \ifthenelse{\boolean{articletitles}}{\emph{{The \lhcb trigger and its performance in 2011}}, }{}\href{https://doi.org/10.1088/1748-0221/8/04/P04022}{JINST \textbf{8} (2013) P04022}, \href{http://arxiv.org/abs/1211.3055}{{\normalfont\ttfamily arXiv:1211.3055}}\relax
\mciteBstWouldAddEndPuncttrue
\mciteSetBstMidEndSepPunct{\mcitedefaultmidpunct}
{\mcitedefaultendpunct}{\mcitedefaultseppunct}\relax
\EndOfBibitem
\bibitem{LHCb-PAPER-2014-047}
LHCb collaboration, R.~Aaij {\em et~al.}, \ifthenelse{\boolean{articletitles}}{\emph{{Precision luminosity measurements at LHCb}}, }{}\href{https://doi.org/10.1088/1748-0221/9/12/P12005}{JINST \textbf{9} (2014) P12005}, \href{http://arxiv.org/abs/1410.0149}{{\normalfont\ttfamily arXiv:1410.0149}}\relax
\mciteBstWouldAddEndPuncttrue
\mciteSetBstMidEndSepPunct{\mcitedefaultmidpunct}
{\mcitedefaultendpunct}{\mcitedefaultseppunct}\relax
\EndOfBibitem
\bibitem{Sjostrand:2007gs}
T.~Sj\"{o}strand, S.~Mrenna, and P.~Skands, \ifthenelse{\boolean{articletitles}}{\emph{{A brief introduction to PYTHIA 8.1}}, }{}\href{https://doi.org/10.1016/j.cpc.2008.01.036}{Comput.\ Phys.\ Commun.\  \textbf{178} (2008) 852}, \href{http://arxiv.org/abs/0710.3820}{{\normalfont\ttfamily arXiv:0710.3820}}\relax
\mciteBstWouldAddEndPuncttrue
\mciteSetBstMidEndSepPunct{\mcitedefaultmidpunct}
{\mcitedefaultendpunct}{\mcitedefaultseppunct}\relax
\EndOfBibitem
\bibitem{LHCb-PROC-2010-056}
I.~Belyaev {\em et~al.}, \ifthenelse{\boolean{articletitles}}{\emph{{Handling of the generation of primary events in Gauss, the LHCb simulation framework}}, }{}\href{https://doi.org/10.1088/1742-6596/331/3/032047}{J.\ Phys.\ Conf.\ Ser.\  \textbf{331} (2011) 032047}\relax
\mciteBstWouldAddEndPuncttrue
\mciteSetBstMidEndSepPunct{\mcitedefaultmidpunct}
{\mcitedefaultendpunct}{\mcitedefaultseppunct}\relax
\EndOfBibitem
\bibitem{Lange:2001uf}
D.~J. Lange, \ifthenelse{\boolean{articletitles}}{\emph{{The EvtGen particle decay simulation package}}, }{}\href{https://doi.org/10.1016/S0168-9002(01)00089-4}{Nucl.\ Instrum.\ Meth.\  \textbf{A462} (2001) 152}\relax
\mciteBstWouldAddEndPuncttrue
\mciteSetBstMidEndSepPunct{\mcitedefaultmidpunct}
{\mcitedefaultendpunct}{\mcitedefaultseppunct}\relax
\EndOfBibitem
\bibitem{davidson2015photos}
N.~Davidson, T.~Przedzinski, and Z.~Was, \ifthenelse{\boolean{articletitles}}{\emph{{PHOTOS interface in C++: Technical and physics documentation}}, }{}\href{https://doi.org/https://doi.org/10.1016/j.cpc.2015.09.013}{Comp.\ Phys.\ Comm.\  \textbf{199} (2016) 86}, \href{http://arxiv.org/abs/1011.0937}{{\normalfont\ttfamily arXiv:1011.0937}}\relax
\mciteBstWouldAddEndPuncttrue
\mciteSetBstMidEndSepPunct{\mcitedefaultmidpunct}
{\mcitedefaultendpunct}{\mcitedefaultseppunct}\relax
\EndOfBibitem
\bibitem{Agostinelli:2002hh}
Geant4 collaboration, S.~Agostinelli {\em et~al.}, \ifthenelse{\boolean{articletitles}}{\emph{{Geant4: A simulation toolkit}}, }{}\href{https://doi.org/10.1016/S0168-9002(03)01368-8}{Nucl.\ Instrum.\ Meth.\  \textbf{A506} (2003) 250}\relax
\mciteBstWouldAddEndPuncttrue
\mciteSetBstMidEndSepPunct{\mcitedefaultmidpunct}
{\mcitedefaultendpunct}{\mcitedefaultseppunct}\relax
\EndOfBibitem
\bibitem{Allison:2006ve}
Geant4 collaboration, J.~Allison {\em et~al.}, \ifthenelse{\boolean{articletitles}}{\emph{{Geant4 developments and applications}}, }{}\href{https://doi.org/10.1109/TNS.2006.869826}{IEEE Trans.\ Nucl.\ Sci.\  \textbf{53} (2006) 270}\relax
\mciteBstWouldAddEndPuncttrue
\mciteSetBstMidEndSepPunct{\mcitedefaultmidpunct}
{\mcitedefaultendpunct}{\mcitedefaultseppunct}\relax
\EndOfBibitem
\bibitem{LHCb-PROC-2011-006}
M.~Clemencic {\em et~al.}, \ifthenelse{\boolean{articletitles}}{\emph{{The \lhcb simulation application, Gauss: Design, evolution and experience}}, }{}\href{https://doi.org/10.1088/1742-6596/331/3/032023}{J.\ Phys.\ Conf.\ Ser.\  \textbf{331} (2011) 032023}\relax
\mciteBstWouldAddEndPuncttrue
\mciteSetBstMidEndSepPunct{\mcitedefaultmidpunct}
{\mcitedefaultendpunct}{\mcitedefaultseppunct}\relax
\EndOfBibitem
\bibitem{Sjostrand:2006za}
T.~Sj\"{o}strand, S.~Mrenna, and P.~Skands, \ifthenelse{\boolean{articletitles}}{\emph{{PYTHIA 6.4 physics and manual}}, }{}\href{https://doi.org/10.1088/1126-6708/2006/05/026}{JHEP \textbf{05} (2006) 026}, \href{http://arxiv.org/abs/hep-ph/0603175}{{\normalfont\ttfamily arXiv:hep-ph/0603175}}\relax
\mciteBstWouldAddEndPuncttrue
\mciteSetBstMidEndSepPunct{\mcitedefaultmidpunct}
{\mcitedefaultendpunct}{\mcitedefaultseppunct}\relax
\EndOfBibitem
\bibitem{Skands:2009zm}
P.~Z. Skands, \ifthenelse{\boolean{articletitles}}{\emph{{The Perugia tunes}}, }{}\href{http://arxiv.org/abs/0905.3418}{{\normalfont\ttfamily arXiv:0905.3418}}\relax
\mciteBstWouldAddEndPuncttrue
\mciteSetBstMidEndSepPunct{\mcitedefaultmidpunct}
{\mcitedefaultendpunct}{\mcitedefaultseppunct}\relax
\EndOfBibitem
\bibitem{LHCb-TDR-009}
LHCb collaboration, \ifthenelse{\boolean{articletitles}}{\emph{{LHCb reoptimized detector design and performance: Technical Design Report}}, }{} \href{http://cdsweb.cern.ch/search?p=CERN-LHCC-2003-030&f=reportnumber&action_search=Search&c=LHCb} {CERN-LHCC-2003-030}, 2003, \url{https://cds.cern.ch/record/630827}\relax
\mciteBstWouldAddEndPuncttrue
\mciteSetBstMidEndSepPunct{\mcitedefaultmidpunct}
{\mcitedefaultendpunct}{\mcitedefaultseppunct}\relax
\EndOfBibitem
\bibitem{LHCb-PUB-2016-021}
L.~Anderlini {\em et~al.}, \ifthenelse{\boolean{articletitles}}{\emph{{The PIDCalib package}}, }{} \href{http://cdsweb.cern.ch/search?p=LHCb-PUB-2016-021&f=reportnumber&action_search=Search&c=LHCb+Notes} {LHCb-PUB-2016-021}, 2016, \url{https://cds.cern.ch/record/2202412}\relax
\mciteBstWouldAddEndPuncttrue
\mciteSetBstMidEndSepPunct{\mcitedefaultmidpunct}
{\mcitedefaultendpunct}{\mcitedefaultseppunct}\relax
\EndOfBibitem
\bibitem{Needham:1082460}
M.~Needham, \ifthenelse{\boolean{articletitles}}{\emph{{Clone track identification using the Kullback--Liebler distance}}, }{} \href{http://cdsweb.cern.ch/search?p=LHCb-2008-002&f=reportnumber&action_search=Search&c=LHCb} {LHCb-2008-002}, 2008, \url{https://cds.cern.ch/record/1082460}\relax
\mciteBstWouldAddEndPuncttrue
\mciteSetBstMidEndSepPunct{\mcitedefaultmidpunct}
{\mcitedefaultendpunct}{\mcitedefaultseppunct}\relax
\EndOfBibitem
\end{mcitethebibliography}

\newpage
\centerline
{\large\bf LHCb collaboration}
\begin
{flushleft}
\small
R.~Aaij$^{38}$\lhcborcid{0000-0003-0533-1952},
A.S.W.~Abdelmotteleb$^{57}$\lhcborcid{0000-0001-7905-0542},
C.~Abellan~Beteta$^{51}$\lhcborcid{0009-0009-0869-6798},
F.~Abudin{\'e}n$^{57}$\lhcborcid{0000-0002-6737-3528},
T.~Ackernley$^{61}$\lhcborcid{0000-0002-5951-3498},
A. A. ~Adefisoye$^{69}$\lhcborcid{0000-0003-2448-1550},
B.~Adeva$^{47}$\lhcborcid{0000-0001-9756-3712},
M.~Adinolfi$^{55}$\lhcborcid{0000-0002-1326-1264},
P.~Adlarson$^{84}$\lhcborcid{0000-0001-6280-3851},
C.~Agapopoulou$^{14}$\lhcborcid{0000-0002-2368-0147}
% ...

{\footnotesize \it

% ...
$^{14}$Universit{\'e} Paris-Saclay, CNRS/IN2P3, IJCLab, Orsay, France\\
% ...
$^{38}$Nikhef National Institute for Subatomic Physics, Amsterdam, Netherlands\\
% ...
$^{47}$Instituto Galego de F{\'\i}sica de Altas Enerx{\'\i}as (IGFAE), Universidade de Santiago de Compostela, Santiago de Compostela, Spain\\
% ...
$^{51}$Physik-Institut, Universit{\"a}t Z{\"u}rich, Z{\"u}rich, Switzerland\\
% ...
$^{55}$H.H. Wills Physics Laboratory, University of Bristol, Bristol, United Kingdom\\
% ...
$^{57}$Department of Physics, University of Warwick, Coventry, United Kingdom\\
% ...
$^{60}$School of Physics and Astronomy, University of Glasgow, Glasgow, United Kingdom\\
$^{61}$Oliver Lodge Laboratory, University of Liverpool, Liverpool, United Kingdom\\
% ...
$^{69}$Syracuse University, Syracuse, NY, United States\\
% ...
$^{84}$Department of Physics and Astronomy, Uppsala University, Uppsala, Sweden, associated to $^{60}$\\
% ...
}
\end{flushleft}
%



\end{document}